\documentclass{iopart}
\usepackage{euscript,iopams,amssymb,amsfonts,graphicx,bm}
\usepackage{pgfplots,setstack}

\usepackage{float}
\usepackage{epsfig}
\usepackage{esint}
\usepackage{chemfig}
\usepackage{cite}

\usepackage{bm,braket}

\eqnobysec

\newcommand{\dis}{\displaystyle}

\renewcommand{\theequation}{\arabic{section}.\arabic{equation}}
\newcommand{\calH}{{\mathcal H}}
\newcommand{\ocalH}{\overline{\mathcal H}}
\newcommand{\calU}{{\mathcal U}}

\newcommand{\calP}{{\mathcal P}}
\newcommand{\calQ}{{\mathcal Q}}
\newcommand{\calS}{{\mathcal S}}

\newcommand{\R}{{\mathbb R}}
\renewcommand{\L}{{\mathbb L}}
\newcommand{\X}{\mathbf{X}}
\renewcommand{\P}{\mathbb{P}}

\newcommand{\x}{\mathbf{x}}

\newcommand{\J}{\widetilde{J}}
\newcommand{\y}{\mathbf{y}}
\renewcommand{\e}{{\mathrm e}}
\newcommand{\E}{{\mathbb E}}

\newcommand{\n}{\mathbf n}

\newcommand{\calT}{{\mathfrak T}}
\newcommand{\calF}{{\mathcal F}}
\renewcommand{\P}{\mathbb P}
\newcommand{\p}{\widetilde{p}}

\newcommand{\f}{\widetilde{f}}
\newcommand{\q}{\widetilde{q}}

\newcommand{\wrho}{\widetilde{\rho}}
\newcommand{\wphi}{\widetilde{\phi}}

\newcommand{\Markov}[2]{\underset{#1}{\overset{#2}{\rightleftharpoons}}}

\newcommand{\of}[1]{\textcolor{black}{#1}}

\renewcommand{\S}{\widetilde{S}}

\begin{document}

 \title[Diffusion-mediated adsorption versus absorption at partially reactive targets]{Diffusion-mediated adsorption versus absorption at partially reactive targets: a renewal approach }


\author{Paul C. Bressloff}
\address{Department of Mathematics, Imperial College London, 
London SW7 2AZ, UK}

\begin{abstract} 
Renewal theory is finding increasing applications in non-equilibrium statistical physics. One example relates the probability density and survival probability of a Brownian particle or an active run-and-tumble particle with stochastic resetting to the corresponding quantities without resetting. A second example is so-called snapping out Brownian motion, which sews together diffusions on either side of an impermeable interface to obtain the corresponding stochastic dynamics across a semi-permeable interface. A third example relates diffusion-mediated surface adsorption-desorption (reversible adsorption) to the case of irreversible adsorption. In this paper we apply renewal theory to diffusion-mediated adsorption processes in which an adsorbed particle may be permanently removed (absorbed) prior to desorption. We construct a pair of renewal equations that relate the probability density and first passage time (FPT) density for absorption to the corresponding quantities for irreversible adsorption. We first consider the example of diffusion in a finite interval with a partially reactive target at one end. We use the renewal equations together with an encounter-based formalism to explore the effects of non-Markovian adsorption/desorption on the moments and long-time behaviour of the FPT density for absorption. We then analyse the corresponding renewal equations for a partially reactive semi-infinite trap and show how the solutions can be expressed in terms of a Neumann series expansion. Finally, we construct higher-dimensional versions of the renewal equations and derive general expression for the FPT density using spectral decompositions. 

\end{abstract}

\maketitle
\section{Introduction}

\begin{figure}[b!]
\centering
\includegraphics[width=12cm]{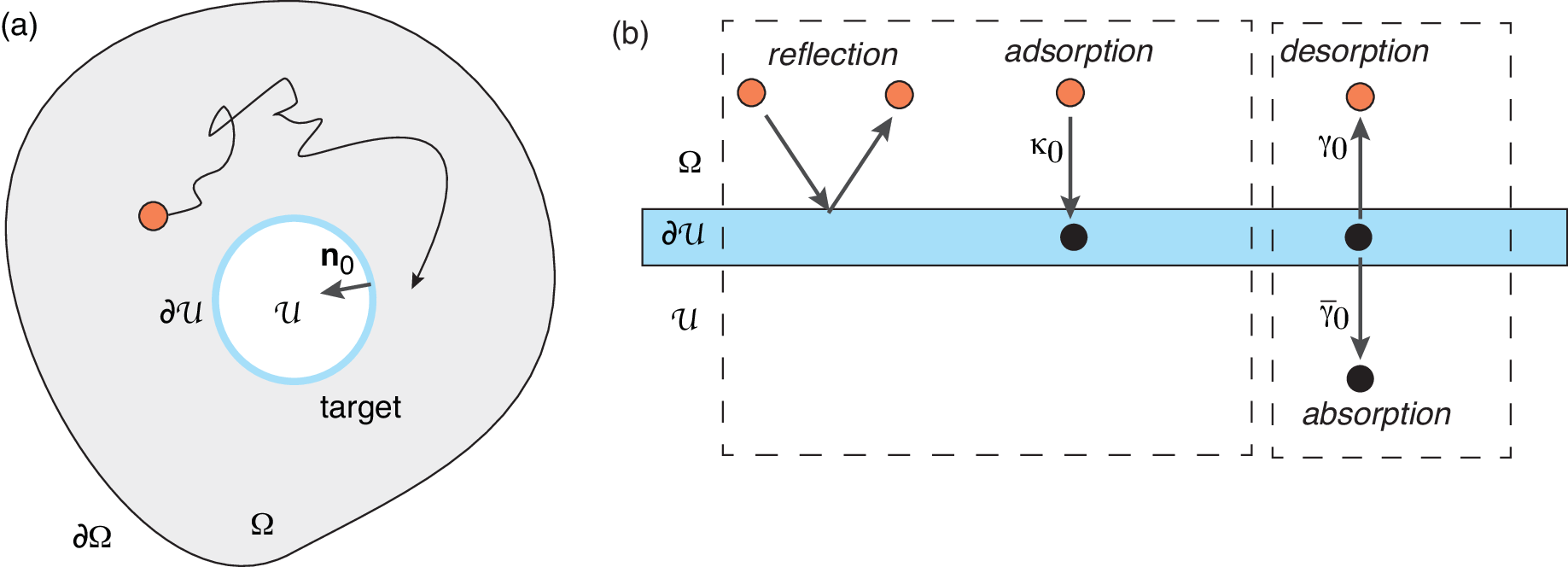} 
\caption{(a) Single-particle diffusion in a bounded domain $\Omega$ containing a partially reactive surface $\partial \calU$ with $\calU\subset \Omega$. The exterior boundary $\partial \Omega$ is totally reflecting. (b)  A particle is either reflected at the boundary $\partial \calU$ or adsorbed at a rate $\kappa_0$. A bound particle either desorbs at a rate $\gamma_0$ or is permanently removed (absorbed) at a rate $\overline{\gamma}_0$.}
\label{fig1}
\end{figure}

\begin{figure}[t!]
\centering
\includegraphics[width=12cm]{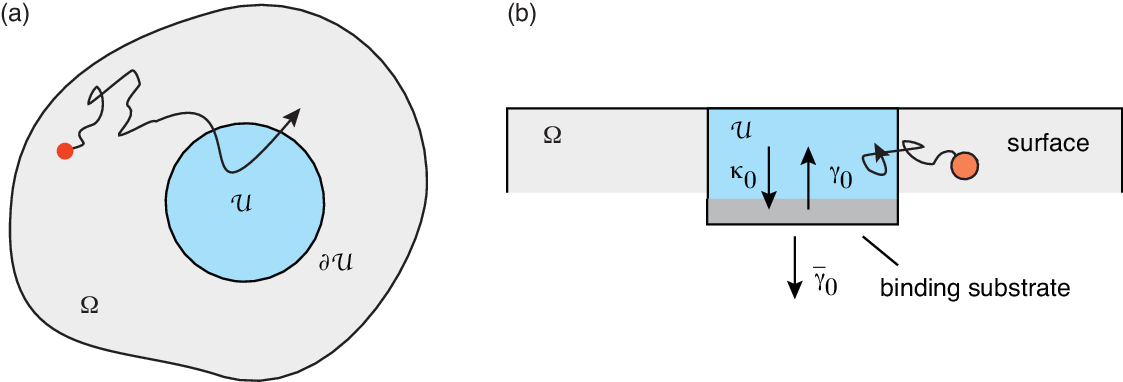} 
\caption{(a) Single-particle diffusion in a bounded domain $\Omega$ containing a partially reactive trap $\calU\subset \Omega$. The exterior boundary $\partial \Omega$ is totally reflecting. (b) Side view. A particle within $\calU$ can be adsorbed at a rate $\kappa_0$, desorb at a rate $\gamma_0$ and be permanently removed (absorbed) at a rate $\overline{\gamma}_0$.}
\label{fig2}
\end{figure}

A classical problem in single-particle diffusion is characterising the first passage
time (FPT) to find a target $\calU\subset \Omega$ within a bounded domain $ \Omega$ \cite{Grebenkov24B}.
At the simplest level, the stochastic dynamics can be modelled as standard Brownian motion, supplemented by the stopping condition that the search process is terminated as soon as the particle reaches the target boundary $\partial \calU$, see Fig. \ref{fig1}(a).  The latter event occurs at the FPT $T=\inf\{t>0, \X(t) \in \partial \calU\}$. However, this scenario oversimplifies what can happen whenever the particle reaches the target. First, a typical surface reaction is unlikely to be instantaneous, instead requiring an alternating sequence of periods of bulk diffusion interspersed with local surface interactions before binding to the surface. In other words, the reactive surface $\partial \calU$ is {\em partially adsorbing}. Second, an adsorbed particle may subsequently unbind and return to diffusion in the bulk ({\em desorption}) or may be permanently transferred to the interior $\calU$ ({\em absorption}). There are thus two levels of partial reactivity, one associated with adsorption from the bulk and the other associated with absorption of the bound or adsorbed particle, see Fig. \ref{fig1}(b). This type of surface reactivity is a common feature of signal transduction in biological cells \cite{Bressloff21B}, where an extracellular signalling molecule (ligand) reversibly binds to the surface membrane. Internalisation of the molecule via an active process known as endocytosis can then trigger a signalling cascade within the cell. In certain applications one finds that the interior $\calU$ rather than the surface $\partial \calU$ acts as a partially reactive target, see Fig. \ref{fig2}(a). In this case a diffusing particle freely enters and exits $\calU$, and while it is diffusing within $\calU$ it binds or adsorbs to the substrate. The particle then either unbinds from the substrate or is permanently removed from the domain $\Omega$, see Fig. \ref{fig2}(b). One important example of this type of process is a protein receptor searching for a synaptic target within the cell membrane of a neuron. Absorption corresponds to active internalisation of the receptor followed by downstream chemical reactions including degradation \cite{Bressloff23a}.

The simplest mathematical implementation of a partially adsorbing boundary $\partial \calU$ is to supplement the diffusion equation for the probability density $p(\x,t)$, $\partial_t p(\x,t)=D{\bm \nabla}^2p(\x,t)$ for all $\x \in \Omega \backslash \calU$, by the Robin boundary condition $D\nabla p(\x,t)\cdot \n_0+\kappa_0 p(\x,t)=0$ for all $\x\in \partial \calU$. (We assume that the exterior boundary $\partial \Omega$ is totally reflecting.) Here $D$ is the diffusivity, $\kappa_0$ is a constant reactivity that specifies the adsorption rate, and $\n_0$ is the unit normal at a point on the boundary $\partial \calU$, see Fig. \ref{fig1}(a). The boundary becomes totally adsorbing  in the limit $\kappa_0\rightarrow \infty$ and totally reflecting when $\kappa_0=0$. (The analog of the Robin BVP for a partially adsorbing target $\calU$ is to take the adsorption rate within $\calU$ to be a constant.) Reversible surface adsorption/desorption processes at the macroscopic level have been studied for many years in physical chemistry, see for example Refs. \cite{Baret68,Adam87,Adam87a,Franses95,Passerone96}. More recently, a comprehensive microscopic theory of such processes has been developed \cite{Reuveni23}. The latter introduces an additional surface probability density $q(\x,t)$ that a particle is bound at the point $\x\in \partial \calU$ following adsorption. Assuming a Robin boundary condition and first order kinetics for desorption, the surface density evolves as $\partial_tq(\x,t)=\kappa_0 p(\x,t) - \gamma_0q(\x,t)$, $\x \in \partial \calU$, where $ \gamma_0$ is a constant rate of desorption. In terms of the underlying stochastic dynamics, both adsorption and desorption are Markov processes. 
Reversible surface adsorption-desorption processes have many features in common with 
reversible recombination–dissociation reactions, which have also been extensively studied \cite{Agmon84,Agmon89,Agmon90,Agmon93,Gopich99,Kim99,Tachiya13,Prustel13,Grebenkov19}. However, the latter are typically based on the Smoluchowski theory of diffusion-limited reactions in which the target is a spherically symmetric molecule or cell in a background of diffusing particles. This contrasts with adsorption-desorption target problems, where diffusion is spatially confined and both $\Omega$ and $\calU$ can have more general shapes. (In applications to physical chemistry, diffusion may be confined between two parallel adsorbing plates.)

Modelling surface adsorption in terms of a constant reactivity $\kappa_0$ is an idealisation of more realistic surface-based reactions \cite{Filoche08}. This has motivated the development of encounter-based models of diffusion-mediated adsorption, which assume that the probability of adsorption depends upon the amount of particle-target contact time \cite{Grebenkov20,Grebenkov22,Bressloff22,Bressloff22a,Grebenkov24}. In the case of a partially adsorbing surface, the amount of contact time is determined by a Brownian functional known as the boundary local time $\ell(t)$ \cite{Ito65,McKean75}. An adsorption event is identified as the first time that the local time crosses a randomly generated threshold $\widehat{\ell}$. This yields the stopping condition $T=\inf\{t>0, \ell(t) >\widehat{\ell} \}$. Different models of adsorption then correspond to different choices of the random threshold probability density $\psi(\ell)$. If $\psi(\ell)$ is an exponential distribution, then the probability of adsorption over an infinitesimal local time increment is independent of the accumulated local time and we have Markovian adsorption. On the other hand, a non-exponential distribution represents non-Markovian adsorption. Physically speaking, the probability of adsorption could depend on some internal state of the adsorbing substrate or particle, and activation/deactivation of this state proceeds progressively by repeated particle-target encounters. The encounter-based formalism can also be applied to a partially absorbing interior trap $\calU$, see Fig. \ref{fig2}(a). The contact time is now given by the Brownian occupation time, which specifies the amount of time the particle spends within $\calU$\cite{Bressloff22,Bressloff22a}.

Recently the encounter-based model of diffusion-mediated surface adsorption has been extended to the case of reversible adsorbing/desorbing surfaces \cite{Grebenkov23}, in which desorption is also taken to be non-Markovian \cite{Evangelista15,Evangelista20}. A key step in the analysis of Ref. \cite{Grebenkov23} is to write down a renewal equation that relates the probability density in the presence of desorption to the corresponding probability density without desorption\footnote{As well as applications to adsorption/desorption processes and semi-permeable membranes, renewal theory provides a powerful method for analysing stochastic processes with resetting \cite{Evans20}. In particular, it allows one to determine the probability density and FPT density with resetting to the corresponding quantities without resetting.}. This is achieved by treating the stochastic dynamics as a sequence of partially reflected Brownian motions in $\Omega\backslash \calU$. Each round is killed when its Brownian local time exceeds a random threshold $\widehat{\ell}$ as in the original formulation of encounter-based adsorption \cite{Grebenkov20}, after which the local time is  reset to zero. A new round is then started after a random waiting time $\tau$ with density $\phi(\tau)$. The Markovian case is recovered by taking both $\psi(\ell)$ and $\phi(\tau)$ to be exponential distributions. The renewal equation is solved using a combination of Laplace transforms and spectral decompositions. Note that an analogous renewal formulation arises in an encounter-based model of diffusion across a semi-permeable interface $\Sigma $ that partitions the search domain $\Omega$ into two parts $\Omega^{\pm}$  \cite{Bressloff22snob,Bressloff23snob}. In this case the stochastic dynamics is formulated in terms of a generalised version of snapping out Brownian motion (BM) \cite{Lejay16}. The latter sews together successive rounds of partially reflecting BMs that are restricted to either $\Omega^+$ or $\Omega^-$. Again, each round is killed when its Brownian local time exceeds a random threshold, after which the local time is reset to zero. A new round is then immediately started in $\Omega_+$ with probability $\sigma_+$ or in $\Omega_-$ with probability $\sigma_-$ such that $\sigma_++\sigma_-=1$.

In this paper we use a renewal approach to analyse the FPT problem for absorption of a particle at a partially reactive target. In particular, we construct a pair of renewal equations that relate both the probability density and FPT density for absorption to the corresponding quantities in the case of irreversible adsorption. The existence of two renewal equations reflects the fact that there are two levels of partial reactivity, one associated with adsorption from the bulk and the other associated with absorption of the bound particle. In order to develop the analysis we initially focus on one-dimensional (1D) diffusion in the finite interval $[0,L]$ with a partially reactive boundary at $x=0$ and a totally reflecting boundary at $x=L$. We first show how the solution to the renewal equations in the Markovian case reproduces the probability density and FPT density obtained by solving a corresponding Robin boundary value problem (BVP), see section 2. We also explore the effects of non-Markovian desorption/absorption on fluctuations of the FPT density for absorption, and discuss possible chemical reaction schemes underlying the non-Markovianity. In section 3 we incorporate non-Markovian adsorption using the encounter-based framework, and show how moments of the FPT density depend on the moments of the random threshold probability density $\psi(\ell)$ (assuming the latter exist). We also derive the long-time asymptotics of the FPT density when $\psi(\ell)$ and $\phi(\tau)$ are heavy-tailed. An example of a 1D partially reactive target is considered in section 4. We take the search domain to be $\Omega=[-L,\infty)$ with a reflecting boundary at $x=-L$ and assume constant rates of adsorption, desorption and absorption within $\calU=[0,\infty)$. We show that even for this relatively simple geometric configuration, the solution of the FPT density for absorption takes the form of an infinite Neumann series expansion of a Fredholm integral equation. Finally, in section 5 we construct higher-dimensional versions of the renewal equations and use these to derive general expressions for the FPT density by extending spectral methods previously developed in Refs. \cite{Grebenkov19a,Grebenkov20,Bressloff22a,Grebenkov23}. We consider both a partially reactive surface $\partial \calU$ and a partially reactive interior $\calU$ as shown in Fig. \ref{fig1}(a) and Fig. \ref{fig2}(a), respectively.

\setcounter{equation}{0}

\section{Diffusion in a finite interval with a partially reactive boundary}

\subsection{The boundary value problem and survival probabilities}

\begin{figure}[b!]
\centering
\includegraphics[width=10cm]{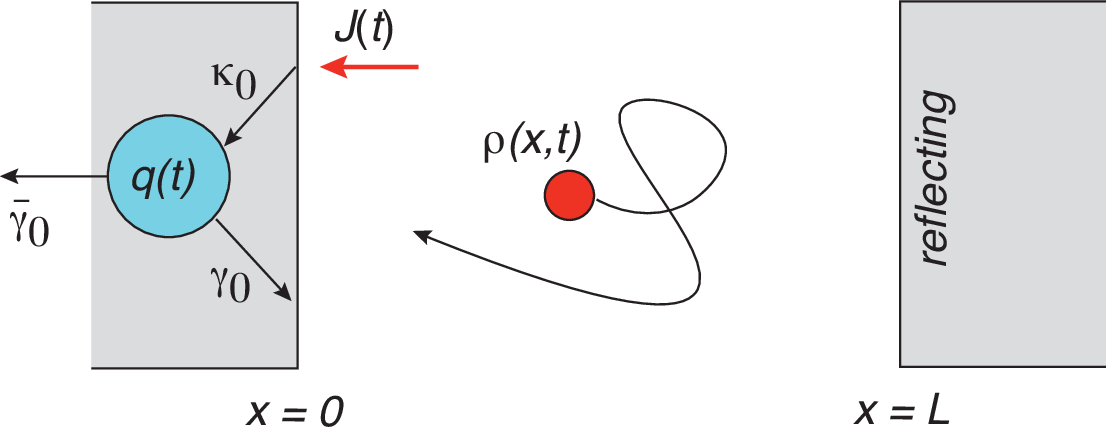} 
\caption{Brownian particle in $[0,L]$ with a partially reactive boundary at $x=0$ and a reflecting boundary at $x=L$. The particle is adsorbed at a rate $\kappa_0$ by binding to the surface and then either desorbs at a rate $\gamma_0$ or is permanently absorbed at a rate $\overline{\gamma}_0$. The probability of being in the bound state is $q(t)$ whereas the probability density of the freely diffusing particle is $\rho(x,t)$. The diffusive flux into the boundary at $x=0$ is $J(t)=D\partial_x\rho(0,t)$.}
\label{fig3}
\end{figure}

Consider a Brownian particle diffusing in the interval $x\in [0,L]$ with a partially reactive boundary or target at $x=0$ and a totally reflecting boundary at $x=L$, see Fig. \ref{fig3}.
(One could view this geometric configuration as the 1D version of diffusion between two concentric spheres $\calU$ and $\Omega$ with radii $R_1$ and $R_2$, respectively, such that $x$ plays the role of the radial coordinate $R-R_1$ and $L=R_2-R_1$. Hence, it is a special case of the search process shown in Fig. \ref{fig1}(a) and analysed in section 5. Alternatively, the particle could be diffusing between two long parallel plates.) We assume that the particle starts in the bulk domain at position $x_0$, so that $\rho(x,0|x_0) = \delta(x - x_0)$ and $q(0)=0$. (If the particle started in the bound state then $\rho(x,0)=0$ for all $x\in [0,L]$ and $q(0)=1$.) The probability density $\rho(x,t|x_0)$ evolves according to the equations
\numparts
\begin{eqnarray}
\label{1Da}
 \frac{\partial \rho(x,t|x_0)}{\partial t}&=D\frac{\partial^2\rho(x,t|x_0)}{\partial x^2} \quad 0<x<L,\\
D\frac{\partial \rho(0,t|x_0)}{\partial x}&=\kappa_0 \rho(0,t|x_0) - \gamma_0 q(t),\quad D\frac{\partial \rho(L,t|x_0)}{\partial x}=0,
\label{1Db}
\end{eqnarray}
with
\begin{equation}
\frac{dq(t)}{dt}=\kappa_0\rho(0,t|x_0) -( \gamma_0+\overline{\gamma}_0)q(t).
\label{1Dc}
\end{equation}
\endnumparts
We interpret the boundary condition at $x=0$ as follows. Whenever the particle hits the boundary (with probability flux $D\partial_x\rho(0,t)$) it either reflects or enters a bound state at a constant rate $\kappa_{0}$. We denote the probability that the particle is in the bound state at time $t$ by $q(t)$. (Note that $q(t)$ also depends on the initial state.) The bound particle then unbinds at a rate $ \gamma_0$ or is permanently removed at a rate $\overline{\gamma}_0$. 

The target at $x=0$ is partially reactive at two levels, see Fig. \ref{fig1}(b). First, there is a non-zero probability of reflecting off the target rather than being {\em adsorbed} by temporarily binding to the target. Second, there is a non-zero probability of unbinding or {\em desorbing} rather than being {\em absorbed} (permanently removed from the system). It follows that we can define two distinct survival probabilities. One specifies the probability that the particle is freely diffusing at time $t$,
\begin{equation}
\label{S}
\calS(x_0,t)=\int_0^{L}\rho(x,t|x_0)dx,
\end{equation}
whereas the other represents the probability that the particle has not yet been absorbed (irrespective of whether it is freely diffusing or bound to the target),
\begin{equation}
\label{Q}
\calQ(x_0,t)=1-\overline{\gamma}_0\int_0^t q(\tau)d\tau.
\end{equation}
Differentiating this pair of equations with respect to time $t$ and using equations (\ref{1Da})-(\ref{1Dc}) shows that
\begin{equation}
\label{dS}
\frac{d\calS(x_0,t)}{dt}=-D\frac{\partial \rho(0,t|x_0)}{\partial x},
\end{equation}
and
\begin{equation}
\label{dQ}
\frac{d\calQ(x_0,t)}{dt}=- \overline{\gamma}_0  q(t).
\end{equation}
That is, $-d\calS/dt$ equals the net flux into the target from the bulk and $-d\calQ/dt$ is the absorption flux (rate of killing). We also note that $\calQ(x_0,0)=1$ for finite $\overline{\gamma}_0$ and
\begin{eqnarray}
&\lim_{t\rightarrow \infty}\calQ(x_0,t)=0=\lim_{t\rightarrow \infty}\calS(x_0,t),\quad \overline{\gamma}_0>0.
\end{eqnarray}
On the other hand, $\calS(x_0,0)$ will depend on the initial conditions. A related point is that, in contrast to $\calQ(x_0,t)$, $\calS(x_0,t)$ may not be a monotonically decreasing function of time $t$. For example, if $q(0)=1$ then $\calS(x_0,0)=0$ and non-monotonicity follows from the fact that $\calS(x_0,t=\infty)=0$, whereas $\calS(x_0,t)>0$ for $0<t<\infty$. Finally, the first passage time (FTP) density for absorption or permanent removal is
\begin{equation}
\calF(x_0,t)=-\frac{d\calQ(x_0,t)}{dt}=\overline{\gamma}_0  q(t).
\end{equation}
The corresponding MFPT 
\begin{equation}
\label{TF}
\calT(x_0)=\int_0^{\infty}t\calF(x_0,t)dt
\end{equation}
can then be written as
\begin{equation}
\calT(x_0)=- \left .\frac{\partial \widetilde{\calF}(x_0,s)}{\partial s}\right |_{s=0}=-\overline{\gamma}_0 \left .\frac{\partial \q(s)}{\partial s}\right |_{s=0}.
\end{equation}
Hence, one way to calculate $\calT(x_0)$ is to solve equations (\ref{1Da})--(\ref{1Dc}) in Laplace space. For the sake of illustration, we assume that the particle is initially at the position $x_0>0$.

Laplace transforming equations (\ref{1Da}) and (\ref{1Db}) gives
\numparts
\begin{eqnarray}
\label{1DLTa}
 & D\frac{\partial^2\wrho(x,s|x_0)}{\partial x^2} -s\wrho(x,s|x_0)= -\delta(x-x_0),\quad x\in (0,L),\\
 &D\frac{\partial \wrho(0,s|x_0)}{\partial x}=\kappa_0\wrho(0,s|x_0)- \gamma_0\widetilde{q}(s),\\
 & D\frac{\partial \wrho(L,s|x_0)}{\partial x}=0,
\label{1DLTb}
\\  &\widetilde{q}(s)=\frac{\kappa_0 \wrho(0,s|x_0)}{s+ \gamma_0+\overline{\gamma}_0}.
\label{1DLTc}
\end{eqnarray}
\endnumparts
The general bounded solution of equation (\ref{1DLTa}) is of the form
\begin{equation}
\wrho(x,s|x_0)=A(s)\cosh(\alpha(s)(L-x))+ G(x,s|x_0),
\end{equation}
where $\alpha(s)=\sqrt{s/D}$ and $G$ is the modified Helmholtz Neumann Green's function satisfying
  \numparts
\begin{eqnarray}
\label{Ga1D}
&D\frac{\partial^2G(x,s|x_0)}{\partial x^2}  -sG(0,s|x_0)  = -\delta(x - x_0), \ 0<x<L,\\ 
&  \partial_xG(0,s|x_0)=0=\partial_xG(L,s|x_0).
 \label{Gb1D}
\end{eqnarray}
\endnumparts
Hence,
\begin{eqnarray}
  \label{G}
\fl   G(x, s| x_0)
 = \frac{\Theta(x_0 - x)g(x, s)\widehat{g}(x_0, s) +\Theta(x - x_0)g(x_0, s)\widehat{g}(x, s)}{\sqrt{sD}\sinh(\sqrt{s/D}L)},
\end{eqnarray}
where $\Theta(x)$ is the Heaviside function and
\begin{eqnarray}
   g(x, s) = \cosh (\sqrt{s/D} x),\quad \widehat{g}(x, s) =\cosh(\sqrt{s/D} (L-x)). 
\end{eqnarray}
Note that
\begin{equation}
\label{G0}
G(0,s|x_0)=\frac{\cosh(\sqrt{s/D} (L-x_0))}{\sqrt{sD}\sinh(\sqrt{s/D} L)}
\end{equation}
and
\begin{eqnarray}
\label{GG1D}
 \lim_{L\rightarrow \infty} G(x, s|x_0) = \frac{1}{2\sqrt{sD}}\left [\e^{-\alpha(s) |x-x_0|}+\e^{-\alpha(s) (x+x_0)}\right ].
 \end{eqnarray}
  The unknown coefficients $A(s)$ is determined from the boundary condition at $x=0$, which yields the explicit solution
\begin{eqnarray}
\label{1Dgir}
\fl  \wrho(x,s|x_0)&= G(x,s|x_0) 
 -\frac{\cosh(\alpha(s)(L-x))}{\cosh(\alpha(s)L)}\frac{\Gamma(s) G(0,s|x_0)}{\Gamma(s)+\alpha(s) \tanh(\alpha(s)L)} , 
\end{eqnarray}
 with
 \begin{equation}
 \label{Gam}
 \Gamma(s)=\frac{1}{D}\frac{\kappa_0[s+\overline{\gamma}_0]}{s+ \gamma_0+\overline{\gamma}_0}.
 \end{equation}
Finally, plugging in the solution for $\wrho$ into equation (\ref{1Dc}) yields
\begin{eqnarray}
\label{LTcalF}
\widetilde{\calF}(x_0,s)&=\frac{\kappa_0\overline{\gamma}_0}{s+ \gamma_0+\overline{\gamma}_0}\wrho(0,s|x_0) \\
&=\frac{\kappa_0\overline{\gamma}_0}{s+ \gamma_0+\overline{\gamma}_0}\frac{\alpha(s)G(0,s|x_0)\tanh(\alpha(s)L)}{\Gamma(s)+\alpha(s) \tanh(\alpha(s)L)}.\nonumber
\end{eqnarray}

 \subsection{Renewal equations}
 
 In this paper we are interested in extending the above model (and its higher-dimensional analogues) to include non-Markovian models of adsorption, desorption and absorption. A powerful method for implementing these extensions is to reformulate the BVP as a renewal equation that separates the diffusion-adsorption process from the desorption-absorption process. Such an approach was recently applied to the case of reversible adsorption-desorption where there is no absorption \cite{Grebenkov23}. Here we develop the corresponding renewal theory when absorption is included.

 First suppose that adsorption is irreversible by setting $ \gamma_0=0$ in equation (\ref{1Db}). (More precisely, we also identify the adsorption and absorption events by taking $\overline{\gamma}_0\rightarrow \infty$.) Denoting the corresponding probability density by $p(x,t|x_0)$ we have the classical Robin BVP
 \numparts
\begin{eqnarray}
\label{Robina}
 \frac{\partial p(x,t|x_0)}{\partial t}&=D\frac{\partial^2p(x,t|x_0)}{\partial x^2} \quad x>0,\\
D\frac{\partial p(0,t|x_0)}{\partial x}&=\kappa_0 p(0,t|x_0) ,\quad D\frac{\partial p(L,t|x_0)}{\partial x}=0,
\label{Robinb}
\end{eqnarray}
\endnumparts
with $p(x,0|x_0)=\delta(x-x_0)$. The survival probability is
\begin{equation}
\label{Q0}
Q(x_0,t)=\int_0^Lp(x,t|x_0)dx,
\end{equation}
and the FPT density for adsorption/absorption is
\begin{equation}
\label{fff}
f(x_0,t)= -\frac{\partial Q(x_0,t)}{\partial t}=\kappa_0 p(0,t|x_0) .
\end{equation}
The last equality follows from differentiating equation (\ref{Q0}) with respect to $t$ and using equations (\ref{Robina}) and (\ref{Robinb}).
The Laplace transformed density $\p(x,s|x_0)$ can be obtained from equation
 (\ref{1Dgir}) by setting $ \gamma_0=0$:
\begin{eqnarray}
\label{1Dp}
 \fl \p(x,s|x_0)= G(x,s|x_0)-\frac{\kappa_0}{D}\frac{\cosh(\alpha(s)(L-x))}{\cosh(\alpha(s)L)}\frac{ G(0,s|x_0)}{\kappa_0/D+\alpha(s) \tanh(\alpha(s)L)}.  
\end{eqnarray}
Similarly,
Laplace transforming equation (\ref{fff}) yields
\numparts
\begin{equation}
\label{1Dfa}
  \f(x_0,s)=\kappa_0\p(0,s|x_0)  
\end{equation}
which, combined with (\ref{1Dp}), becomes
\begin{equation}
\label{1Dfb}
  \f(x_0,s) =\frac{\kappa_0\alpha(s) G(0,s|x_0)\tanh(\alpha(s)L)}{\kappa_0/D+\alpha(s) \tanh(\alpha(s)L)}.
\end{equation}
\endnumparts

Now suppose that  $0< \gamma_0,\overline{\gamma}_0<\infty$ and the particle binds to the target (is adsorbed) at a time $t_a$, say. Let $\sigma(\tau)$ be the probability that it is still bound at time $t_a+\tau$. (The probability $\sigma(\tau)$ is distinct from the probability $q(t)$ appearing in the BVP (\ref{1Da})-(\ref{1Dc}), since the former is conditioned on adsorption occurring at time $\tau=0$.) We have
 \begin{equation}
 \frac{d\sigma}{d\tau}=-\gamma \sigma(\tau),\quad \sigma(0)=1,\quad  \gamma= \gamma_0+\overline{\gamma}_0,
 \end{equation}
such that $\sigma(\tau)=\e^{- \gamma \tau}$ for $\tau\geq 0.$
Let $\phi(\tau)$ denote the waiting time density for either desorption or absorption to occur. We can then set $d\sigma /d\tau =-\phi(\tau)$ with
 \begin{equation}
 \label{LTphi}
\phi(\tau)= \gamma \e^{-\gamma \tau},\quad  \wphi(s)=\frac{ \gamma_0+\overline{\gamma}_0}{s+ \gamma_0+\overline{\gamma}_0}.
\end{equation}
In addition, denoting the splitting probabilities for desorption and absorption by $\pi_d$ and $\pi_b$, respectively, we have
 \begin{equation}
 \pi_d =\frac{ \gamma_0}{\gamma_0+\overline{\gamma}_0},\quad \pi_b=\frac{\overline{\gamma}_0}{ \gamma_0+\overline{\gamma}_0}.
 \end{equation}
The densities $\rho(x,t|x_0)$ and $\calF(x_0,t)$ in the presence of desorption-absorption are then related to the corresponding densities $p(x,t|x_0)$ and $f(x_0,t)$ according to the following pair of first renewal equations:
\numparts
\begin{eqnarray}
\label{ren1}
 \fl \rho(x,t|x_0)&=p(x,t|x_0)+\pi_d\int_0^td\tau' \int_{\tau'}^t d\tau\,  \rho(x,t-\tau|0)  \phi(\tau-\tau') f(x_0,\tau'),\\
 \fl \calF(x_0,t)&=\pi_b\int_0^td\tau  {\phi}(t-\tau) f(x_0,\tau)+\pi_d \int_0^td\tau' \int_{\tau'}^t d\tau\, \calF(0,t-\tau) \phi(\tau-\tau') f(x_0,\tau').\nonumber\\
 \fl
 \label{ren2}
 \end{eqnarray}
 \endnumparts
The first term on the right-hand side of the renewal equation (\ref{ren1}) for the density $\rho(x,t|x_0)$ represents the contribution from all sample paths that have not been adsorbed over the interval $[0,t]$, which is determined by the solution to the Robin BVP (\ref{Robina}) and (\ref{Robinb}). On the other hand, the second term represents all sample paths that are first adsorbed at a time $\tau' $ with probability $f(x_0,\tau')d\tau'$, remain in the bound state until desorbing at time $\tau$ with probability $\pi_d \phi(\tau-\tau')d\tau$, after which the particle may bind an arbitrary number of times before reaching $x$ at time $t$. Turning to the renewal equation (\ref{ren2}) for the FPT  $\calF(x_0,t)$, the first term on the right-hand side represents all sample paths that are first adsorbed at time $\tau'$ and are subsequently absorbed at time $t$ without desorbing, which occurs with probability $\pi_b{\phi}(t-\tau')f(x_0,\tau')d\tau'dt$. In a complementary fashion, the second term sums over all sample paths that are first adsorbed at time $\tau'$, desorb at time $\tau$ and are ultimately absorbed at time $t$ following an arbitrary number of additional adsorption events. 

The renewal equations can be solved using Laplace transforms and the convolution theorem.
Equation (\ref{ren1}) becomes
\begin{equation}
\wrho(x,s|x_0)=\p(x,s|x_0)+ \wrho(x,s|0)\pi_d \wphi(s) \f(x_0,s).
 \end{equation}
 Setting $x_0=0$ and rearranging shows that
 \begin{equation}
 \wrho(x,s|0)=\frac{\p(x,s|0)}{1-\pi_d \wphi(s) \f(0,s)}
 \end{equation}
 and, hence,
 \begin{equation}
\label{LTren1}
\wrho(x,s|x_0)=\p(x,s|x_0)+ \Lambda(x_0,s)\p(x,s|0)
 \end{equation}
with
\begin{equation}
\label{lambo}
\Lambda(x_0,s)=\frac{\pi_d \wphi(s) \f(x_0,s)}{1-\pi_d \wphi(s) \f(0,s)}.
\end{equation}
Similarly, Laplace transforming the second renewal equation (\ref{ren2}) gives
 \begin{equation}
 \widetilde{\calF}(x_0,s)=\wphi(s)\bigg [\pi_b \f(x_0,s)+\pi_d \widetilde{\calF}(0,s) \f(x_0,s)\bigg ].
 \end{equation}
 Setting $x_0=0$ and rearranging implies that
 \begin{equation}
  \widetilde{\calF}(0,s)=\frac{\pi_b\wphi(s) \f(0,s)}{1-\pi_d\wphi(s) \f(0,s)}
  \end{equation}
 and thus
  \begin{eqnarray}
 \widetilde{\calF}(x_0,s)&= \pi_b \wphi(s)\bigg [ \f(x_0,s)+\Lambda(x_0,s)\f(0,s)\bigg ]\nonumber \\
 &=\frac{\pi_b \wphi(s) \f(x_0,s)}{1-\pi_d \wphi(s) \f(0,s)}.
  \label{LTren2}
 \end{eqnarray}
 Also note that substituting equation (\ref{1Dfa}) into the first line of (\ref{LTren2}) implies that 
  \begin{eqnarray}
 \widetilde{\calF}(x_0,s)&=\kappa_0\pi_b\wphi(s)\bigg [ \p(0,s|x_0)+\Lambda(x_0,s)\p(0,s|0)\bigg ]\nonumber \\
 &=\kappa_0\pi_b   \wphi(s)\wrho(0,s|x_0).
\label{LTren2a}
 \end{eqnarray}
 In the case of the exponential waiting time densities and their Laplace transforms (\ref{LTphi}), it can be shown after some algebra that equation (\ref{LTren1}) with $\p(x,s|x_0)$ and $\f(x_0,s)$ given by equations (\ref{1Dp}) and (\ref{1Dfb}), respectively, recovers the solution (\ref{1Dgir}) for $\wrho(x,s|x_0)$ of the Robin BVP. Note, in particular that
 \begin{equation}
 \label{lambo2}
   \Lambda(x_0,s)= \frac{\kappa_0\pi_d \wphi(s)\alpha(s)G(0,s|x_0)\tanh(\alpha(s)L)}{\Gamma(s)+\alpha(s) \tanh(\alpha(s)L)} .
\end{equation}
The corresponding solution (\ref{LTcalF}) for the FPT density then follows from equation (\ref{LTren2a}). 

\subsection{Non-exponential waiting time densities and extended kinetic schemes}

One immediate generalisation of the renewal equations (\ref{ren1}) and (\ref{ren2}) is to take the waiting time kernel $\phi(\tau)$ to be a non-exponential function of $\tau$, see also Ref. \cite{Grebenkov23}. Desorption and absorption are no longer Markov processes, at least at the level of first order kinetics. A possible mechanism for generating a non-exponential waiting time density $\phi(\tau)$ is an extended kinetic scheme that involves transitions between multiple bound states.
Suppose that, following surface adsorption, the bound particle has to undergo a sequence of reversible reactions before either being absorbed or desorbed. Let $S_0$ denote the free particle and $S_m$, $m=1,\ldots,M$, one of the bound states. Consider the reaction scheme

\begin{equation}
\schemestart
       \subscheme{$S_0$}
     \arrow(a--c){->[$\kappa(t)$]}
        \subscheme{$S_1$}
     \arrow(2a--2c){<=>[$\beta_1$][$\alpha_1$]}
      \subscheme{$S_2$}
     \ldots
      \arrow(2a--2c){<=>[$\beta_{M-1}$][$\alpha_{M-1}$]}
     \subscheme{$S_M$}
     \arrow(cd--a)[-45]
    \subscheme{$S_0$}
   \arrow(@cd--b)[45]
   \subscheme{$\emptyset$}   
    \schemestop
    \label{scheme1}
    \end{equation}
    
 \noindent where $\kappa(t)=\kappa_0\rho(0,t|x_0)$. Let $q_{m}(t)$ denote the probability that the particle is in bound state $m$. The first equation in (\ref{1Db}) becomes
 \numparts
\begin{eqnarray}
\label{kina}
D\frac{\partial \rho(0,t|x_0)}{\partial x}&=\kappa_0 \rho(0,t|x_0) - \gamma_0 q_M(t), 
\label{kinb}
\end{eqnarray}
while (\ref{1Dc}) takes the more general form
\begin{eqnarray}
\fl & \frac{dq_1(t)}{dt}=\kappa_0\rho(0,t|x_0) - \beta_1q_1(t)+\alpha_1 q_2(t)
\label{kinc}\\
\fl &	\frac{dq_{m}}{dt}= \beta _{m-1} q_{m-1}-(\alpha_{m-1}+\beta_{m})q_{m}+\alpha_{m}q_{m+1},\quad 1<m<M\\
\fl &	\frac{dq_{M}}{dt}= \beta _{M-1} q_{M-1}-(\alpha_{M-1}+\gamma_0+\overline{\gamma}_0)q_{M}.
\label{kine}
\end{eqnarray}
\endnumparts
Laplace transforming equations (\ref{kinc})--(\ref{kine}), assuming that the particle is initially unbound, gives
 \numparts 
\begin{eqnarray}
\label{kineticLTa}
	&s\q_{m}(s) = \widetilde{\kappa}(s) \delta_{m,1} +\sum_{m'=1}^M\Gamma_{mm'}(0,\gamma)\q_{m'}(s)
	\label{kineticLTb}
\end{eqnarray}
\endnumparts
for $m=1,\ldots,M$,
where $\gamma=\overline{\gamma}_0+\gamma_0$ and $\Gamma_{mm'}(a,b)$ is an element of the tridiagonal matrix
\begin{equation}
\fl {\bm \Gamma}(a,b)=\left (\begin{array}{cccccc} -a -\beta_1 & \alpha_1 &0 &0 &\ldots&0 \\ \beta_1&-\alpha_1-\beta_2 & \alpha_2 &0 & \ldots&0
\\0& \beta_2 &-\alpha_2-\beta_3 & \alpha_3&\ldots&0\\
\vdots & \vdots & \vdots & \vdots &\vdots  &\vdots \\
0 & 0 &\ldots & 0 & \beta_{M-1} & -\alpha_{M-1} -b
\end{array} \right ).
\end{equation}
It follows that
\begin{equation}
\label{Cm}
\q_{m}(s)=  (s{\bf I}-{\bm \Gamma}(0, \gamma)^{-1}_{m1}\widetilde{\kappa}(s).
\end{equation}
Comparing with our renewal formulation, we can make the identification
\begin{equation}
   \widetilde{\phi}(s)=\gamma \q_M(s),\quad \q_M(s)= \bigg [s{\bf I}-{\bm \Gamma}(0, \gamma)\bigg ]^{-1}_{M1}.
\end{equation}

An alternative kinetic scheme is to assume that desorption occurs from the initial bound state $S_1$ but one or more reversible reactions have to occur before absorption can occr:
\begin{equation}
S_0\Markov{\gamma_0} {\kappa(t)}S_1 \Markov{\alpha_1}{\beta_1}S_2\cdots  \Markov{\alpha_{M-1}}{\beta_{M-1}}S_M\overset{\overline{\gamma}_0}\rightarrow \emptyset.
\end{equation}
Equation (\ref{1Db}) remains unchanged whereas (\ref{1Dc}) becomes
\begin{eqnarray}
\fl & \frac{dq_1(t)}{dt}=\kappa_0\rho(0,t|x_0) - (\beta_1+\gamma_0)q_1(t)+\alpha_1 q_2(t),
\label{kinc2}\\
\fl &	\frac{dq_{m}}{dt}= \beta _{m-1} q_{m-1}-(\alpha_{m-1}+\beta_{m})q_{m}+\alpha_{m}q_{m+1},\quad 1<m<M\\
\fl &	\frac{dq_{M}}{dt}= \beta _{M-1} q_{M-1}-(\alpha_{M-1}+\overline{\gamma}_0)q_{M}.
\label{kine2}
\end{eqnarray}
It follows that
\begin{equation}
\label{Cm2}
\q_{m}(s)=  (s{\bf I}-{\bm \Gamma}(\gamma_0,\overline{\gamma}_0)^{-1}_{m1}\widetilde{\kappa}(s),
\end{equation}
and we have two different delay kernels for desorption and absorption. That is, in the renewal equations (\ref{ren1}) and (\ref{ren2}) we replace $\pi_d\phi(\tau)$ and $\pi_b \phi(\tau)$ by the kernels $\phi_d(\tau)$ and $\phi_b(\tau)$, respectively, where
\begin{equation}
\widetilde{\phi}_d(s)= \gamma_0   \bigg [s{\bf I}-{\bm \Gamma}(\gamma_0,\overline{\gamma}_0)^{-1}_{11}\bigg ]
\end{equation}
and
\begin{equation}
\widetilde{\phi}_b(s) = \overline{\gamma}_0   \bigg [s{\bf I}-{\bm \Gamma}(\gamma_0,\overline{\gamma}_0)^{-1}_{M1}\bigg ].
\end{equation}
For simplicity, we will only consider the first reaction scheme (\ref{scheme1}).

 \subsection{Moments of the FPT density for absorption}

Equation (\ref{LTren2}) can be used to express each moment of the FPT density $\widetilde{\calF}(x_0,s)$ (assuming they exist) in terms of moments of $\f(x_0,s)$. We use the fact that $\widetilde{\calF}(x_0,s)$ and $\f(x_0,s)$ are moment generating functions. That is,
\numparts
 \begin{equation}
 T_{n}(x_0):=\int_0^{\infty}t^n f(x_0,t)dt=\left . \left (-\frac{d}{ds}\right )^n \f(x_0,s)\right |_{s=0},
 \end{equation}
 and
 \begin{equation}
 \calT_{n}(x_0):=\int_0^{\infty}t^n \calF(x_0,t)dt=\left . \left (-\frac{d}{ds}\right )^n \widetilde{\calF}(x_0,s)\right |_{s=0}.
 \end{equation}
 \endnumparts
 In particular, $\calT=\calT_1$ and $T=T_1$ are the MFPTs. The moment equations can be obtained either by taking the $n$th order derivative of equation (\ref{LTren2}) and then setting $s\rightarrow 0$, or by Taylor expanding the right-hand side of equation (\ref{LTren2}) in powers of $s$ and identifying the sum of $O(s^n)$ terms with the $n$th moment multiplied by $(-1)^n/n!$. Here we use the latter method to derive the first two moments. In the following we assume that $\phi(\tau)$ has finite  moments. (Heavy-tailed waiting time densities will be considered in section 3.) Substituting the series expansions
 \numparts
 \begin{eqnarray}
  \f(x_0,s)&\sim 1-sT (x_0)+s^2T_2(x_0)/2+O(s^3),\\ \wphi(s)&\sim 1-s\langle \tau\rangle +s^2\langle \tau^2\rangle/2+O(s^3)
  \end{eqnarray}
 \endnumparts
into equation (\ref{LTren2}), we obtain the approximation
\begin{eqnarray}
\label{big}
  \fl & \widetilde{\calF} (x_0,s) \sim  \pi_b\bigg [
   \frac{1-s\sigma_1(x_0)+s^2 \sigma_2(x_0)/2}{1-\pi_d[1-s\sigma_1(0)+s^2 \sigma_2(0)/2]}\bigg ]   \\
  \fl &\sim \bigg [1-s\sigma_1(x_0)+s^2 \sigma_2(x_0)/2\bigg ] 
 \bigg [1-\frac{\pi_d}{\pi_b}\bigg (s\sigma_1(0)-s^2\sigma_2(0)/2\bigg )+\left (\frac{\pi_d}{\pi_b}\right )^2s^2 \sigma_1(0)^2\bigg ],\nonumber
\end{eqnarray}
with
\begin{equation}
 \sigma_1 (x_0)=T (x_0)+\langle \tau \rangle,\ \sigma_2(x_0)=T_2(x_0)+\langle \tau^2\rangle + 2\langle \tau\rangle T(x_0),
\end{equation}

First consider the MFPT. Collecting the $O(s)$ terms on the right-hand side of equation (\ref{big}) yields
 \begin{eqnarray}
 \calT(x_0)=T(x_0)+  \langle \tau \rangle+\frac{\pi_d}{\pi_b}\bigg [T(0)+ \langle \tau \rangle\bigg ].
\label{MFPT}
 \end{eqnarray}
The result (\ref{MFPT}) has a simple physical interpretation. Irrespective of the number of desorption events, the particle takes a mean time $T(x_0)$ to be adsorbed for the first time starting from $x_0$, and takes a mean time $\langle \tau \rangle$ to be absorbed following the final adsorption event. The probability of exactly $n$ desorption events is $p_n=\pi_b\pi_d^n$ with
\begin{equation}
\sum_{n=0}^{\infty}p_n= \pi_b\sum_{n=0}^{\infty}\pi_d^n=\frac{\pi_b}{1-\pi_d}=1.
\end{equation}
The mean number of such excursions is 
\begin{eqnarray}
\overline{n}=\sum_{n=0}^{\infty}np_n=\frac{\pi_b\pi_d}{(1-\pi_d)^2}=\frac{\pi_d}{\pi_b},
\end{eqnarray}
and the mean time between excursions is $T_d =\sigma_1(0)=T (0)+\langle \tau \rangle$.
Hence, equation (\ref{MFPT}) can be rewritten as
\begin{equation}
\calT (x_0)=T (x_0)+\overline{n}\, T_d+\langle \tau \rangle.
\end{equation}
Finally, using equation (\ref{1Dfb}), we can calculate the MFPT $T(x_0)$ in the absence of desorption to give
\begin{eqnarray}
\label{calT0}
T(x_0)=  \frac{L^2}{2D}-\frac{(L-x_0)^2}{2D}+\frac{L}{\kappa_0} .
\end{eqnarray}
The first two terms on the right-hand represent the classical MFPT for a totally adsorbing boundary at $x=0$ without desorption and under the initial position $x_0$. The third term accounts for the additional time accrued when the boundary is partially absorbing with constant reactivity $\kappa_0$. The MFPT $ \calT(x_0)$ is clearly a monotonically increasing function of $x_0$, $\pi_d$, the mean waiting time $\langle \tau\rangle=\mu/\gamma$ and the mean adsorption time $1/\kappa_0$. If we fix $\langle \tau\rangle$, then the MFPT is the same irrespective of the choice of waiting time density. That is, in order to distinguish between exponential and non-exponential densities $\phi(\tau)$ we need to consider higher-order moments.

\begin{figure}[t!]
\centering
\includegraphics[width=9.5cm]{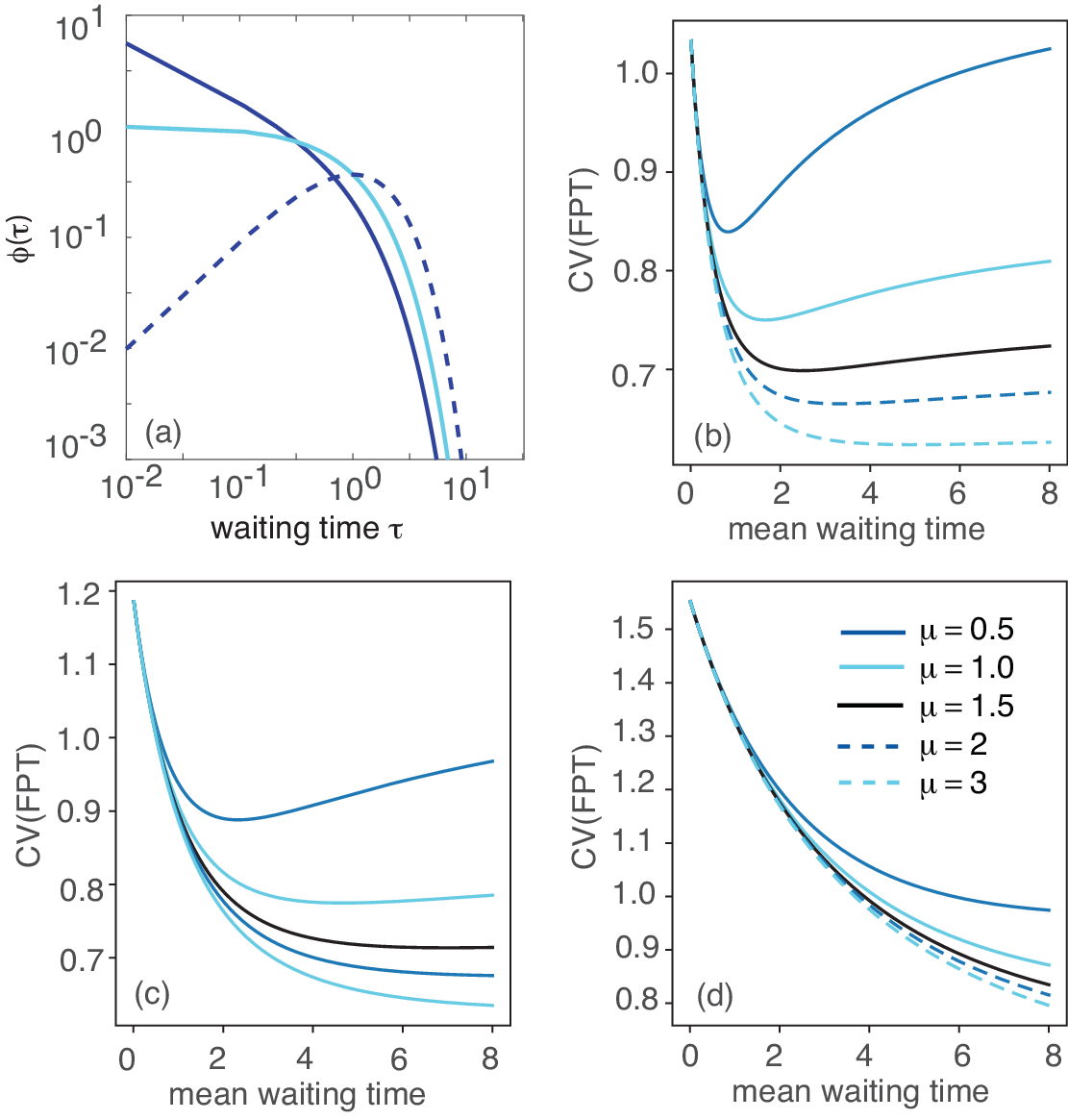} 
\caption{Markovian adsorption with non-Markovian desorption/absorption. (a) Plots of the waiting time density $\phi(\tau)$ for the gamma distribution with $\gamma=1$ and various values of $\mu$. (b-d) Plots of coefficient of variation (CV) of the FPT density for absorption as a function of $\langle \tau \rangle $ and various values of the parameter $\mu$ of the gamma distribution (\ref{phigam}). (b) $L=1$. (c) $L =2$. (d) $L=5$. Other parameters are $D=1$, $\kappa_0=1$, $\pi_d=\pi_b=0.5$ and $x_0=0$.}
\label{fig4}
\end{figure}

Collecting the $O(s^2)$ terms on the right-hand side of equation (\ref{big}) yields the second-order FPT moment
  \begin{eqnarray}
 \fl \calT_2(x_0)&=T_2(x_0)+\langle \tau^2\rangle + 2\langle \tau\rangle T(x_0) 
  +\frac{\pi_d}{\pi_b}\bigg [T_2(0)+\langle \tau^2\rangle + 2\langle \tau\rangle T(0)\bigg ]  
  \nonumber \\
\fl &\quad   +\frac{\pi_d}{\pi_b}\bigg [T(x_0)+ \langle \tau \rangle\bigg ]\bigg [T(0)+ \langle \tau \rangle\bigg ] +2\left (\frac{\pi_d}{\pi_b}\right )^2\bigg [T(0)+ \langle \tau \rangle\bigg ]^2,
\label{MFPT2}
 \end{eqnarray}
 with $T_2(x_0)$ obtained from equation (\ref{1Dfb}):
 \begin{eqnarray}
\frac{T_2(x_0)}{2}&=  \frac{5L^4+(L-x_0)^4 -6L^2(L-x_0)^2}{4!D^2}
  \nonumber \\
  &\quad +\frac{[L^2-(L-x_0)^2]L}{\kappa_0D} +\frac{L^2}{\kappa_0^2}+\frac{L^3 }{3\kappa_0D}.
 \label{calT2}
\end{eqnarray}
One well-known example of a non-exponential waiting time density with finite moments is the gamma distribution, see Fig. \ref{fig4}(a):
\begin{equation}
\label{phigam}
\phi(\tau)=\frac{\gamma(\gamma \tau)^{\mu-1}\e^{-\gamma_0 \tau}}{\Gamma(\mu)},\quad \mu >0,
\end{equation}
where $\Gamma(\mu)$ is the gamma function. If $\mu=1$, then we recover the exponential distribution $\phi(\tau)=\gamma\e^{-\gamma \tau}$. It can be seen from Fig. \ref{fig4}(a) that the probability of small values of the waiting time $\tau$ can be decreased relative to an exponential distribution by taking $\mu >1$. This could represent a bound state that is initially relatively stable, but becomes more unstable  as $\tau$ increases. On the other hand, the probability of small values of $\tau$  is increased when $\mu < 1$ so that the bound state is initially more unstable.
The corresponding Laplace transform is
\begin{equation}
\widetilde{\phi}(s)=\left (\frac{\gamma}{s+\gamma}\right )^{\mu},
\end{equation}
and the waiting time moments are
\begin{equation}
\langle \tau^n\rangle =\left .\left (-\frac{d}{ds}\right )^n\wphi(s)\right |_{s=0}=\frac{\mu (\mu+1)\ldots (\mu+n-1)}{\gamma^{n}}.
\end{equation}

For the sake of illustration, we focus on the coefficient of variation (CV) which is defined according to
\begin{equation}
CV=\frac{\sqrt{ \calT_2(x_0)- \calT(x_0)^2}}{ \calT(x_0)}.
\end{equation}
The CV is a measure of the relative size of fluctuations about the MFPT. In Fig. \ref{fig4}(b-d). we plot the CV as a function of the mean waiting time $\langle \tau\rangle $ in the case of the gamma distribution (\ref{phigam}) for different values of $\mu$ and the length $L$. The dependence on $\mu$ arises from the identity $\langle \tau^2\rangle = (1+1/\mu)\langle \tau\rangle^2$. It can be seen that the choice of local time threshold density $\psi$ for a given MFPT has a strong effect on the size of fluctuations, particularly when $\tau >\sqrt{L^2/D}$.

 \section{Encounter-based model of non-Markovian adsorption}
 
 In the case of a constant rates of adsorption, desorption and absorption, the renewal equations (\ref{ren1}) and (\ref{ren2}) provide an alternative representation of the stochastic process evolving according to equations (\ref{1Da}) and (\ref{1Db}), in which there is an explicit summation over the number of desorption events occur prior to absorption. One advantage of the renewal approach is that it is straightforward to incorporate non-Markovian model of absorption and desorption without needing to specify a particular chemical reaction scheme. Moreover, in Laplace space it provides an efficient way of determining the FPT for absorption if the corresponding FPT for adsorption is already known. As previously highlighted in the case of reversible adsorption \cite{Grebenkov23}, yet another advantage of the renewal approach is that it can be extended to include an encounter-based model of non-Markovian adsorption. In this section we develop the corresponding theory in the presence of absorption.
    
\subsection{Irreversible adsorption.}

\begin{figure}[b!]
\centering
\includegraphics[width=8cm]{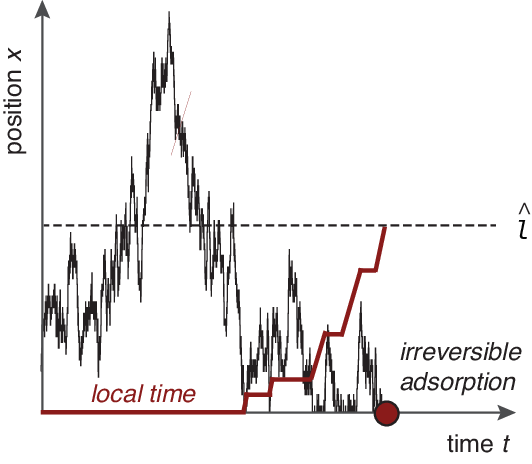} 
\caption{Encounter-based model of irreversible adsorption. Example trajectory of a Brownian particle with a single-level partially reactive boundary at $x=0$. When the local time $\ell(t)$ (shown in red) crosses a randomly generated threshold $\widehat{l}$, the particle is permanently adsorbed.}
\label{fig5}
\end{figure}

Suppose, for the moment, that adsorption is irreversible ($ \gamma_0=0$) so that the particle is permanently removed from the bulk domain at the first adsorption event, see Fig. \ref{fig5}. As mentioned in the introduction, encounter-based models of diffusion-mediated surface adsorption assume that the probability of adsorption depends upon the boundary local time $\ell(t)$ \cite{Grebenkov20,Grebenkov22,Bressloff22,Bressloff22a,Grebenkov24}. In the case of a target at $x=0$, the local time is defined according to
\begin{equation}
\ell(t)=\lim_{\epsilon \rightarrow 0}\frac{D}{\epsilon}\int_0^t\Theta(\epsilon -X(\tau))d\tau.
\end{equation}
(The factor of D means that $\ell(t)$ has units of length.) It can
be proven that $\ell(t)$ exists and is a nondecreasing, continuous function of $t$  \cite{Ito65,McKean75}.
The particle is adsorbed at $x=0$ at the stopping time
 \begin{equation}
\label{exp}
{\mathcal T}=\inf\{t>0:\ \ell(t)>\widehat{\ell}\},\quad \P[\widehat{\ell}>\ell]\equiv \Psi(\ell) .
\end{equation}
Since $\ell(t)$ is a nondecreasing process, the condition $t < {\mathcal T}$ is equivalent to the condition $\ell(t)<\widehat{\ell}$. There are several complementary approaches to constructing the solution of an encounter-based model for a general threshold distribution $\Psi$, including integral/spectral methods \cite{Grebenkov20}, Feynman-Kac formulae \cite{Bressloff22}, and weak representations based on empirical measures and It\^o's formula \cite{Bressloff24}. The third version defines the local time propagator as
\begin{equation}
P(x,\ell,t|x_0)=\bigg \langle \delta(x-X(t))\delta(\ell-\ell(t))\bigg \rangle_{X(0)=x_0},
\end{equation}
where $\langle\cdot \rangle$ denotes the expectation with respect to all sample paths $X(t)$ that satisfy the Skorokhod stochastic differential equation (SDE)
\begin{equation}
dX(t)=\sqrt{2D}dW(t)+d\ell(t)-d\overline{\ell}(t),\quad X(0)=x_0,
\end{equation}
with $W(t)$ a Wiener process and 
\begin{equation}
\widehat{\ell}(t)=\lim_{\epsilon \rightarrow 0}\frac{D}{\epsilon}\int_0^t\Theta(\epsilon-|L-X(\tau)|)d\tau.
\end{equation}
The differentials $d\ell(t)=D\delta(X(t)dt$ and $d\widehat{\ell}(t)=D\delta(L-X(t))$ are impulses that keep the particle within the domain $[0,L]$. (We do not explcitly track $\widehat{\ell}(t)$ in the local time propagator, since the boundary $x=L$ is totally reflecting.)

Introducing the pair of densities
\numparts
\begin{eqnarray}
\fl  p^{\Psi}(x,t|x_0)&= \bigg \langle \delta(x-X(t)\Psi(L(t))\bigg \rangle_{X(0)=x_0}
=\int_0^{\infty}\Psi(\ell) P(x,\ell,t|x_0)d\ell, \\
\fl \nu^{\psi}(x,t|x_0)&=\bigg \langle \delta(x-X(t)\psi(L(t))\bigg \rangle_{X(0)=x_0}  =\int_0^{\infty}\psi(\ell) P(x,\ell,t|x_0)d\ell,
\end{eqnarray}
\endnumparts
with $\psi(\ell) =-\Psi'(\ell)$ the probability density for the local time threshold,
it can be proved that $p^{\Psi}$ and $\nu^{\Psi}$ are related according to the equations \cite{Grebenkov20,Bressloff24}
\numparts
\begin{eqnarray} 
\label{pPsia}
 \frac{\partial p^{\Psi}(x,t|x_0)}{\partial t} 
&=D\frac{\partial^2 p^{\Psi}(x,t|x_0)}{\partial x^2} ,\ x>0,\\
\left . D\frac{\partial p^{\Psi}(x,t|x_0) }{\partial x}\right |_{x=0}&=D\nu^{\psi}(0,t|x_0),\\
  \left . D\frac{\partial p^{\Psi}(x,t|x_0) }{\partial x}\right |_{x=L}&=0.
\label{pPsib}
\end{eqnarray}
\endnumparts
For a general local time threshold distribution $\Psi$, we do not have a closed equation for the marginal density $p^{\Psi}(x,t|x_0)$. However, in the particular case of the exponential distribution $\Psi(\ell)=\e^{-\kappa_0\ell /D}$, we have $\psi(\ell)=\kappa_0 \Psi(\ell)/D$ so that equations (\ref{pPsia}) and (\ref{pPsib}) reduce to the classical Robin BVP given by equations  (\ref{Robina}) and (\ref{Robinb}).
Hence, the solution of the Robin BVP is equivalent to the Laplace transform of the local time propagator with respect to $\ell$ \cite{Grebenkov20,Grebenkov22,Bressloff22,Bressloff22a}:
\begin{equation}
\calP(x,z,t|x_0)\equiv \int_0^{\infty} \e^{-z \ell} P(x,\ell,t|x_0)d\ell=p(x,t|x_0)_{\kappa_0=zD}.
\end{equation}
Assuming that the Laplace transform $\calP(x,z,t)$ can be inverted with respect to $z$, the solution for a general distribution $\Psi$ is
 \begin{eqnarray}
  \label{poo}
  p^{\Psi}(x,t|x_0) &= \int_0^{\infty} \Psi(\ell){\mathbb L}_{\ell}^{-1}\calP(x,z,t|x_0)d\ell.
  \end{eqnarray}
  The corresponding FPT density is
  \begin{eqnarray}
  \label{foo}
  f^{\Psi}(x_0,t):&=D\nu^{\psi}(0,t|x_0)  
 = D\int_0^{\infty} \psi(\ell){\mathbb L}_{\ell}^{-1}\calP(0,z,t|x_0)d\ell .
  \end{eqnarray}

We can calculate $\p^{\Psi}(x,s|x_0)$ and $ \f^{\Psi}(x_0,s)$ by Laplace transforming equation (\ref{poo}) with respect to $t$ and setting $\calP(x,z,t|x_0) =\p(x,s|x_0)_{\kappa_0=zD} $ for $\p(x,s|x_0)$ given by equation (\ref{1Dp}). This yields
\begin{eqnarray}
\fl \calP(x,z,s|x_0)&=G(x,s|x_0) 
  -\frac{\cosh(\alpha(s)(L-x))}{\cosh(\alpha(s)L)}\frac{z G(0,s|x_0)}{z+\alpha(s) \tanh(\alpha(s)L)}\nonumber
\end{eqnarray}
so that
\begin{eqnarray}
\fl & {\mathbb L}^{-1} [\calP(x,z,s|x_0)] =\left [G(x,s|x_0)-G(0,s|x_0)\frac{\cosh(\alpha(s)(L-x))}{\cosh(\alpha(s)L)}\right ]\delta_{\ell,0} \\
 \fl &\quad +\alpha(s)\tanh(\alpha(s)L)\frac{\cosh(\alpha(s)(L-x))}{\cosh(\alpha(s)L)}\exp\bigg(-\ell \alpha(s)\tanh(\alpha(s)L)\bigg )G(0,s|x_0).\nonumber
 \label{inv1}
\end{eqnarray}
In particular,
\begin{eqnarray}
\label{inv2}
\fl  & {\mathbb L}^{-1} [\calP(0,z,s|x_0)]=\alpha(s)\tanh(\alpha(s)L) \exp\bigg(-\ell \alpha(s)\tanh(\alpha(s)L)\bigg )G(0,s|x_0). 
\end{eqnarray}
Finally, the Laplace transformed equations (\ref{poo}) and (\ref{foo}) show that
\numparts
\begin{eqnarray}
\fl &\p^{\Psi}(x,s|x_0) 
=\left [G(x,s|x_0)-G(0,s|x_0)\frac{\cosh(\alpha(s)(L-x))}{\cosh(\alpha(s)L)}\right ]\\
 \fl&\quad +\alpha(s)\tanh(\alpha(s)L)\frac{\cosh(\alpha(s)(L-x))}{\cosh(\alpha(s)L)}\widetilde{\Psi}\bigg(\ell \alpha(s)\tanh(\alpha(s)L)\bigg )G(0,s|x_0)\nonumber ,\\
 \fl  \f^{\Psi}(x_0,s)&=   \frac{\cosh(\sqrt{s /D}(L-x_0))}{\cosh(\sqrt{s/D}L)}\widetilde{\psi}\bigg( \alpha(s)\tanh(\alpha(s)L)\bigg ).
 \label{fPsi}
\end{eqnarray}
\endnumparts

\begin{figure}[t!]
\centering
\includegraphics[width=12cm]{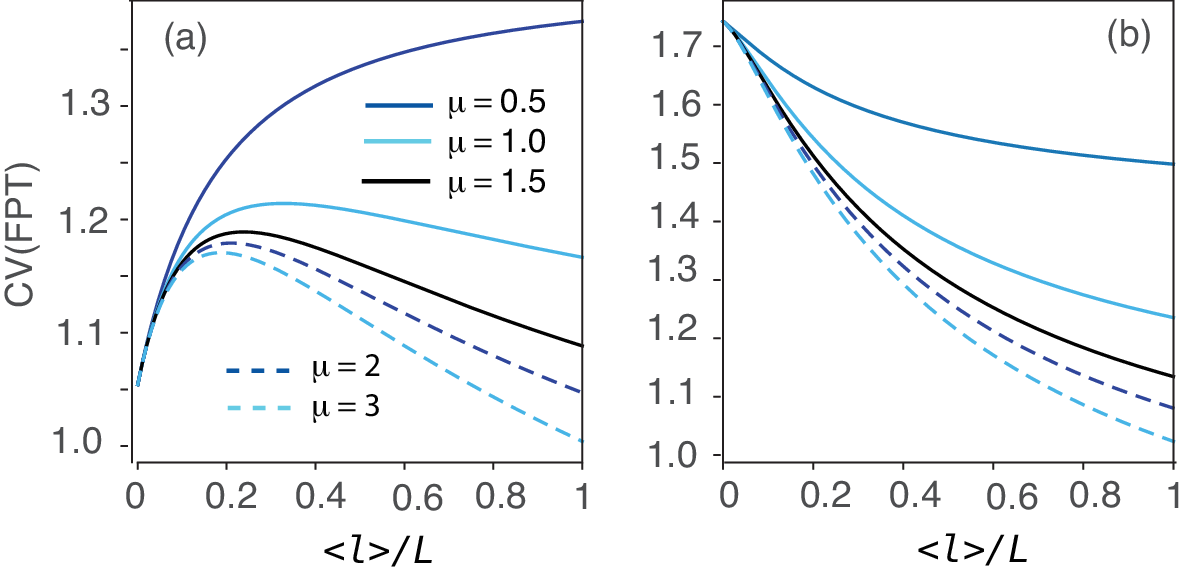} 
\caption{Irreversible non-Markovian adsorption. Coefficient of variation (CV) of the FPT density $\f^{\Psi}(x_0,t)$ of equation (\ref{foo}). The CV is plotted as a function of the mean local time threshold $\langle\ell \rangle $ and various values of the parameter $\mu$ for $\psi(\ell)$ given by the gamma distribution (\ref{phigam}). Other parameters are $D=1$ and (a) $x_0=L/2$. (b) $x_0=L/5$.}
\label{fig6}
\end{figure}

 Suppose, for the moment, that $\psi(\ell)$ has finite first moments and the asymptotic expansion
 \begin{equation}
 \widetilde{\psi}(z)\sim 1-z\langle \ell \rangle+z^2\langle \ell^2\rangle/2,
\end{equation}
 where $\langle \ell\rangle=-\widetilde{\psi}'(0)$ is the mean local time threshold etc.
 Substituting into equation (\ref{fPsi}) and performing a Taylor series expansion in $s$, we find that the MFPT and second-order moment without desorption are 
\numparts
 \begin{equation}
 T^{\Psi}(x_0)=\frac{L^2}{2D}-\frac{(L-x_0)^2}{2D}
+\frac{\langle \ell\rangle  L}{D},\label{TPsi1}
\end{equation}
\begin{eqnarray}
  \fl \frac{T_2^{\Psi}}{2}&=\frac{5L^4+(L-x_0)^4 -6L^2(L-x_0)^2}{4!D^2} 
   +\frac{[L^2-(L-x_0)^2]L\E[\ell]}{D^2} 
 +\frac{\E[\ell^2]L^2}{2D^2}+\frac{L^3\E[\ell]}{3D^2}.\nonumber \\
 \fl 
 \label{TPsi2}
 \end{eqnarray}
 \endnumparts
 Analogous to the case of Markovian adsorption and non-Markovian desorption/absorption, see section  2.3, we introduce the CV of the FPT density
 \begin{equation}
CV^{\Psi}=\frac{\sqrt{ T_2^{\Psi}(x_0)- T^{\Psi}(x_0)^2}}{ T^{\Psi}(x_0)},
\end{equation}
and take $\psi(\ell)$ to be the gamma distribution (\ref{phigam}) with $\tau \rightarrow \ell$.
We can then explore how the CV depends on the local time threshold density $\psi(\ell)$  by plotting the CV as a function of $\langle \ell \rangle$ for different values of $\mu$. Example plots are shown in Fig. \ref{fig6}, which illustrate the dependence of the CV on the choice of density $\psi(\ell)$.

 \subsection{Renewal equations}

\begin{figure}[t!]
\centering
\includegraphics[width=12cm]{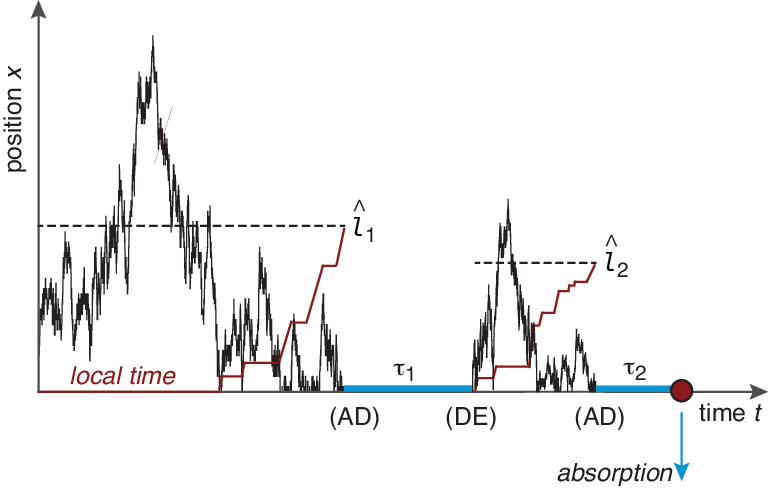} 
\caption{Encounter-based model of partial adsorption combined with desorption and absorption. Example trajectory of a Brownian particle with a two-level partially reactive boundary at $x=0$. When the local time $\ell(t)$ (shown in red) crosses a randomly generated threshold $\widehat{l}_1$, it is adsorbed (AD) and the local time is reset to zero. After a waiting time $\tau_1$, the particle desorbs (DE) and the local time starts increasing again until crossing a new random threshold $\widehat{\ell}_2$. Following a second waiting time $\tau_2$, the particle is permanently absorbed. }
\label{fig7}
\end{figure}

 Now suppose that, following an adsorption event, the particle enters a bound state and the local time resets to zero. The particle remains in the bound state for a random waiting time $\tau$ generated from a density $\phi(\tau)$, after which the particle either desorbs and diffuses until another adsorption event occurs or is permanently killed (absorbed). An example trajectory is illustrated in Fig. \ref{fig7}. 
The assumption that the local time is reset following each adsorption event means that it is not straightforward to write down a PDE for the  the generalised probability density $\rho^{\Psi}(x,t|x_0)$ and FPT density $\calF^{\Psi}(x_0,t)$ in the presence of desorption/absorption.
One of the powerful features of the renewal approach is that  we can immediately write down an encounter-based version of equations (\ref{ren1}) and (\ref{ren2}):
\numparts
\begin{eqnarray}
\label{EBren1}
\fl  \rho^{\Psi}(x,t|x_0)&=p^{\Psi}(x,t|x_0)+\pi_d\int_0^td\tau' \int_{\tau'}^t d\tau\,  \rho^{\Psi}(x,t-\tau|0)\phi(\tau-\tau') f^{\Psi}(x_0,\tau'),\\
\label{EBren2}
\fl  \calF^{\Psi}(x_0,t)\  &=\pi_b\int_0^td\tau \phi(t-\tau') f^{\Psi}(x_0,\tau') \nonumber \\
\fl &\quad +\pi_d\int_0^td\tau' \int_{\tau'}^t d\tau\, \calF^{\Psi}(0,t-\tau) \phi(\tau-\tau') f^{\Psi}(x_0,\tau').
 \end{eqnarray}
 \endnumparts
Note that in the special case of reversible adsorption/desorption ($\pi_b = 0$), $\calF^{\Psi}$ no longer exists and equation (\ref{EBren1}) is equivalent to the renewal equation introduced in Ref. \cite{Grebenkov23}. The latter is obtained by taking $\int_0^{\infty}\phi(\tau)d\tau =1$ and iterating the Volterra-type integral equation. That is,
\begin{eqnarray}
 \fl &\rho^{\Psi}(x,t|x_0) 
 =p^{\Psi}(x,t|x_0)+\int_0^td\tau' \int_{\tau'}^t d\tau\,  p^{\Psi}(x,t-\tau|0)\phi(\tau-\tau') f^{\Psi}(x_0,\tau')\nonumber \\
 \fl &\quad +\int_0^td\tau'' \int_{\tau''}^t d\tau' \int_{\tau'}^t d\tau\,  p^{\Psi}(x,t-\tau|0)\phi(\tau-\tau') f^{\Psi}(0,\tau')\phi(\tau'-\tau'') f^{\Psi}(x_0,\tau'')+\ldots \nonumber 
 \end{eqnarray}
The second term on the right-hand side represents sample paths involving exactly one adsorption/desorption event, the third term represents sample paths involving exactly two adsorption/desorption events etc. In Laplace space equations (\ref{EBren1}) and (\ref{EBren2}) become 
 \begin{equation}
\label{EBLTren1}
\wrho^{\Psi}(x,s|x_0)=\p^{\Psi}(x,s|x_0)+ \Lambda^{\Psi}(x_0,s)\p^{\Psi}(x,s|0)
 \end{equation}
and
\begin{eqnarray}
\fl  \widetilde{\calF}^{\Psi}(x_0,s)&=\pi_b\wphi(s)\bigg [ \f^{\Psi}(x_0,s)+\Lambda^{\Psi}(x_0,s)\f^{\Psi}(0,s)\bigg ]=\frac{\pi_b\wphi(s) \f^{\Psi}(x_0,s)}{1-\pi_d\wphi(s) \f^{\Psi}(0,s)} ,
  \label{EBLTren2}
 \end{eqnarray}
 with $\f^{\Psi}(x_0,s)$ given by equation (\ref{fPsi}) and
\begin{equation}
 \Lambda^{\Psi}(x_0,s)=\frac{\pi_d\wphi(s) \f^{\Psi}(x_0,s)}{1-\pi_d\wphi(s) \f^{\Psi}(0,s)}.
\end{equation}
(In contrast to Markovian adsorption, there is no simple relationship between $\f^{\Psi}$ and $\p^{\Psi}$.)

Suppose that both $\phi(\tau)$ and $\psi(\ell)$ have finite first moments and consider  the asymptotic expansions
 \numparts
 \begin{eqnarray}
 & \f^{\Psi}(x_0,s)\sim 1-sT^{\Psi} (x_0)+s^2T^{\psi}_2(x_0)/2+O(s^3),\\ &\wphi(s)\sim 1-s\langle \tau\rangle +s^2\langle \tau^2\rangle/2+O(s^3)
 \end{eqnarray}
 \endnumparts
 with $T^{\Psi} (x_0)$ and $T^{\psi}_2(x_0)$ given by equations (\ref{TPsi1}) and (\ref{TPsi2}), respectively. Using identical arguments to the derivation of equation (\ref{MFPT}), we obtain the MFPT relation
 \begin{eqnarray}
  \calT^{\Psi}(x_0)&\equiv -\left. \frac{\partial \calF^{\Psi}(x_0,s)}{\partial s}\right |_{s=0}\nonumber \\
  &=T^{\Psi}(x_0)+\frac{\pi_d}{\pi_b}\bigg [T^{\Psi}(0)+\langle \tau \rangle\bigg ]+ \langle \tau \rangle.
\label{EBMFPT}
 \end{eqnarray}
Combining equations (\ref{TPsi1}) and (\ref{EBMFPT}) yields the result
  \begin{eqnarray}
  \fl \calT^{\Psi}(x_0)&=\frac{L^2}{2D}-\frac{(L-x_0)^2}{2D}+
\frac{\langle \ell\rangle L}{D}  
 +\frac{\pi_d}{\pi_b}\left [\frac{\langle \ell\rangle L}{D}+\langle \tau \rangle\right ]+ \langle \tau \rangle.
 \label{TPpsiphi}
 \end{eqnarray}
 An analogous result holds for the second-order FPT moment.

\subsection{Asymptotics of the FPT density for heavy-tailed distributions $\psi(\ell)$ and $\phi(\tau)$} 
\begin{figure}[b!]
\centering
\includegraphics[width=12.5cm]{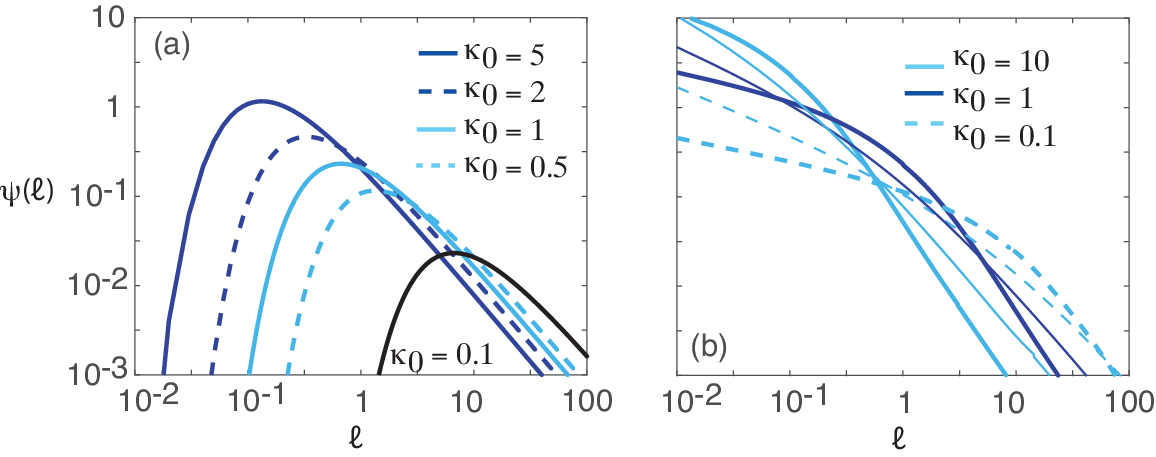} 
\caption{Two examples of a heavy-tailed distribution $\psi(\ell)$. (a) One-side Levy distribution for different values of $\kappa$. (b) Mittag-Leffler distribution for different values of $\kappa$ with $\mu=0.5 $ (thin curves) and $\mu=0.75$ (thick curves).}
\label{fig8}
\end{figure}

We could now explore how the CV of the FPT density depends on the choice of the joint densities $\psi(\ell)$ and $\phi(\tau)$ by combining equations (\ref{TPsi1}), (\ref{TPsi2}) and (\ref{TPpsiphi}) etc.. Instead, here we consider what happens when at least one of the densities is heavy-tailed. For the sake of illustration, suppose that it is the local time threshold density $\psi(\ell)$.
The Taylor expansion of $\f^{\Psi}(x_0,s)$ about $s=0$ in powers of $s$ then breaks down and $T^{\Psi}(x_0)$ etc. no longer exists. However, it is still possible to perform a small-$s$ expansion for specific choices of $\widetilde{\psi}(s)$ in order to characterise the long-time behaviour of the FPT density. For the sake of illustration, we consider two different examples of heavy-tailed distributions as illustrated in Fig. \ref{fig8}. (For a more extensive list, see Ref. \cite{Grebenkov20}.)

The first example is the one-sided L\'evy-Smirnov distribution
 \begin{equation}
 \psi_{\rm ls}(\ell)=\kappa_0 \frac{\e^{-1/(\kappa_0 \ell)}}{\sqrt{\pi} (\kappa_0 \ell)^{3/2}},\quad \widetilde{\psi}_{\rm ls}(z)=\e^{-2\sqrt{z/\kappa_0}}.
 \end{equation}
 This could represent an adsorbing surface that has an optimal range of reactivity \cite{Grebenkov20}. Substituting for $\widetilde{\psi}_{\rm ls}$ into equation (\ref{fPsi}b) gives
 \begin{eqnarray}
 \fl  \f^{\Psi}(x_0,s)  
= \frac{\cosh(\sqrt{s /D}(L-x_0))}{\cosh(\sqrt{s/D}L)}\exp\left (-2\sqrt{\frac{ \alpha(s)\tanh(\alpha(s)L)}{\kappa_0}}\right ). 
 \end{eqnarray}
 For small $s$, we have the approximation
 \begin{eqnarray}
    \f^{\Psi}(x_0,s)& \sim \left (1-s \frac{L^2-(L-x_0)^2}{2K}\right )\left (1-2\sqrt{\frac{s L}{\kappa_0 D}}\right )\nonumber \\
    &= 1-2\sqrt{\frac{s L}{\kappa_0 D}}+ h.o.t.
 \end{eqnarray}
 Hence, the leading-order large-$t$ approximation of $f^{\Psi}(x_0,t)$ is
 \begin{equation}
 f^{\Psi}(x_0,t)\sim \frac{2}{|\Gamma(-1/2)|t^{1/2+1}}\sqrt{\frac{L}{\kappa_0 D}},\quad t \rightarrow \infty .
 \end{equation}
The square-root dependence on $t$ is a consequence of the heavy-tailed L\'evy distribution $\psi_{\rm ls}(\ell)$ that determines adsorption at $x=0$. Note that the short-term contribution to the FPT density is negligible due to inactivity of the boundary for small thresholds $\widehat{\ell}$, see Fig. \ref{fig8}(a). 
The corresponding asymptotics of the FPT with desorption is then obtained as follows. Assuming that $\phi(\tau)$ has finite moments and keeping only leading order terms in powers of $\sqrt{s}$ we have 
\begin{eqnarray}
\fl  \widetilde{\calF}^{\Psi}(x_0,s) 
 \sim \pi_b \frac{1-2\sqrt{\frac{s L}{\kappa_0 D}}}{1-\pi_d\left (1-2\sqrt{\frac{s L}{\kappa_0 D}}\right )}\sim 1-\frac{2}{\pi_b}\sqrt{\frac{s L}{\kappa_0 D}}. 
\end{eqnarray}
Hence, $\pi_b{\calF}^{\Psi}(x_0,t)$ has the same leading-order large-$t$ approximation as $f^{\Psi}(x_0,t)$. 

The second example is the Mittag-Leffler distribution
\begin{equation}
 \psi_{\rm ml}(\ell)=-E_{\mu,0}(-(\kappa_0 \ell)^{\mu})/\ell, \quad E_{\mu,0}(z)=\sum_{k=0}^{\infty} \frac{z^k}{\Gamma(\mu k)}
 \end{equation}
for $0<\mu<1$. The corresponding Laplace transform is
\begin{equation}
 \widetilde{\psi}_{\rm ml}(z)=\frac{\kappa_0^{\mu}}{\kappa_0^{\mu}+z^{\mu}}.
 \end{equation}
 Substituting for $\widetilde{\psi}_{\rm ml}$ in equation (\ref{fPsi}b) now gives
 \begin{eqnarray}
 \fl  \f^{\Psi}(x_0,s) 
  = \frac{\cosh(\sqrt{s /D}(L-x_0))}{\cosh(\sqrt{s/D}L)}\frac{\kappa_0^{\mu}}{\kappa_0^{\mu}+(\alpha(s)\tanh(\alpha(s)L)^{\mu}}.  
 \end{eqnarray}
 For small $s$, we have the approximation
 \begin{eqnarray}
 \f^{\Psi}(x_0,s)&\sim \left (1-s \frac{L^2-(L-x_0)^2}{2D}\right )\left (1-\left (\frac{s L}{\kappa_0 D}\right )^{\mu}\right )\nonumber \\
  &= 1-s \frac{L^2-(L-x_0)^2}{2 D}-\left (\frac{sL}{\kappa_0D}\right )^{\mu}+ h.o.t.
 \end{eqnarray}
 Since $0<\mu < 1$, it follows that the $s^{ \mu}$ term dominates for small $s$.
Hence, the leading-order large-$t$ approximation of $f^{\Psi}(x_0,t)$ is
 \begin{equation}
 f^{\Psi}(x_0,t)\sim \frac{1}{t^{ \mu+1}|\Gamma(-\mu)|}\left (\frac{ L}{\kappa_0D}\right )^{\mu},\quad t \rightarrow \infty .
 \end{equation}
 Again we find that $\pi_b{\calF}^{\Psi}(x_0,t)$ has the same leading-order large-$t$ approximation as $f^{\Psi}(x_0,t)$.

 Finally, we consider the leading order behaviour when both $\phi(\tau)$ and $\psi(\ell)$ are heavy-tailed such that
 \begin{equation}
 \widetilde{\psi}(z )\sim 1- \ell_{0}^{\alpha}z^{\alpha},\quad \wphi(s)\sim 1-\tau_{0}^{\beta} s^{\beta},\quad  0<\alpha,\beta < 1,
 \end{equation}
 where $\ell_0$ and $\tau_0$ set the length and time scales, respectively.
Substituting into equations  (\ref{fPsi})and  (\ref{EBLTren2}) gives
\begin{eqnarray}
 \widetilde{\calF}^{\Psi}(x_0,s) 
& \sim \pi_b \frac{1-\left({s \ell_0L}{/ D}\right )^{\alpha}-(\tau_0s)^{\beta}}{1-\pi_d\left (1- \left({s \ell_0L}{/ D}\right )^{\alpha}- (\tau_0s)^{\beta}\right )}\nonumber \\
 &\sim 1-\frac{1}{\pi_b}\bigg [\left({s \ell_0L}{/ D}\right )^{\alpha}+(\tau_0s)^{\beta}\bigg ].
\end{eqnarray}
There are then three possibilities as $ t \rightarrow \infty$:
 \begin{eqnarray}
 \calF^{\Psi}(x_0,t)\sim  \left \{  \begin{array}{cc} \frac{\dis 1}{\dis |\Gamma(-\alpha)|}\frac{\dis 1}{\dis t^{\alpha+1}}\left (\frac{\dis \ell_0 L}{\dis D}\right )^{\alpha},& \alpha <\beta,\\ \\
   \frac{\dis 1}{\dis |\Gamma(-\beta)|}\frac{\dis \tau_0^{\beta}}{\dis t^{\beta+1}}, & \alpha >\beta\\ \\
    \frac{\dis 1}{\dis |\Gamma(-\alpha)|}\frac{\dis 1}{\dis t^{\alpha+1}}\left (\frac{\dis \ell_0 L}{\dis D}+\tau_0\right )^{\alpha},& \alpha =\beta
 \end{array}
 \right . .
 \end{eqnarray}
 Note that a similar asymptotic analysis was carried out in Ref. \cite{Grebenkov23} for reversible adsorption/desorption. In this case, the quantities of interest are the steady-state
 probability distributions for being in the unbound (diffusing) or bound (adsorbed) states. The asymptotic analysis then provides details regarding the long-time relaxation to the steady state.

\section{Semi-infinite partially reactive target}

\begin{figure}[b!]
\centering
\includegraphics[width=10cm]{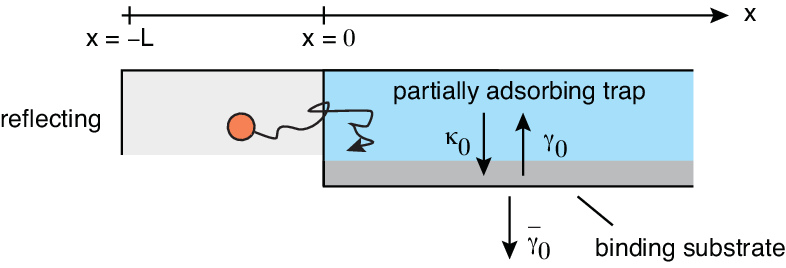} 
\caption{1D diffusion in the domain $\Omega=[-L,\infty)$ with a semi-infinite partially reactive trap $\calU=[0,\infty)$. The boundary $x=-L$ is totally reflecting. The particle within $\calU$ can be adsorbed at a rate $\kappa_0$, desorb at a rate $\gamma_0$ and be permanently removed (absorbed) at a rate $\overline{\gamma}_0$.}
\label{fig9}
\end{figure}

The analysis of partially reactive targets $\calU$ is considerably more involved even in the case of a 1D target with irreversible adsorption \cite{Bressloff22a}. In order to illustrate this, consider a semi-infinite target $\calU=[0,\infty)$ and  $\Omega =[-L,\infty)$, see Fig. \ref{fig9}. For convenience, we set $\rho(x,t|x_0)=\rho_1(x,t|x_0)$ for $x\in[-L,0]$ and $\rho(x,t|x_0)=\rho_2(x,t|x_0) $ for $x \geq 0$. The 1D analogue of the BVP (\ref{1Da}) -- (\ref{1Dc}) is
\numparts
\begin{eqnarray}
\label{1DBVPa}
 \fl &\frac{\partial \rho_1(x,t|x_0)}{\partial t} =D\frac{\partial^2 \rho_1(x,t|x_0)}{\partial x^2},\ -L<x<0,  \\
 \fl & \frac{\partial \rho_1(x,t|x_0)}{\partial x}=0,\quad x=-L,\nonumber \\
 \fl &\frac{\partial \rho_2(x,t|x_0)}{\partial t}=D\frac{\partial^2 \rho_2(x,t|x_0)}{\partial x^2}- \kappa_0 \rho_2(x,t|x_0)+\gamma_0 q(x,t|x_0),\, x>0,\\
\fl &\frac{\partial q(x,t|x_0)}{\partial t}=\kappa_0\rho_2(x,t|x_0) -( \gamma_0+\overline{\gamma}_0)q(x,t|x_0).
\label{1DBVPc}
\end{eqnarray}
Here $q(x,t|x_0)$ is the probability density that the particle is bound at a point $x\in [0,\infty)$.
We also have matching conditions at the interface $x=0$:
\begin{eqnarray}
  \rho_1(0,t|x_0)&=\rho_2(0,t|x_0),\ \frac{\partial \rho_1}{\partial x}(0,t|x_0)  = \frac{\partial\rho_2}{\partial x}(0,t|x_0).
  \label{1DBVPd}
\end{eqnarray}
\endnumparts
The corresponding BVP in Laplace space is
\numparts
\begin{eqnarray}
\label{trapBVPa}
\fl  & D\frac{\partial^2 \rho_1(x,s|x_0)}{\partial x^2}-s\rho(x,s|x_0) =0,\quad  -L<x<0,  \\
\fl  & \frac{\partial \rho_1(x,s|x_0)}{\partial x}=0,\quad x=-L,
 \label{trapBVPb}\\
 \fl  &D\frac{\partial^2 \rho_2(x,s|x_0)}{\partial x^2}-(s+\kappa(s))\rho_2(x,s|x_0)=- \delta(x-x_0),\quad  x,x_0>0.
  \label{trapBVPc}
  \end{eqnarray}
  where
  \begin{equation}
  \kappa(s)=\kappa_0 \frac{s+\overline{\gamma}_0}{s+\gamma_0+\overline{\gamma}_0},
  \end{equation}
together with the matching conditions
\begin{eqnarray}
  \rho_1(0,s|x_0)&=\rho_2(0,s|x_0),\ \frac{\partial \rho_1}{\partial x}(0,s|x_0)  = \frac{\partial\rho_2}{\partial x}(0,s|x_0).
  \label{trapBVPd}
\end{eqnarray}
\endnumparts 
Note that the Laplace transform of equation (\ref{1DBVPc}) implies that
\begin{equation}
\q(x,s|x_0)=\frac{\kappa_0\wrho_2(x,s|x_0)}{s+\gamma_0+\overline{\gamma}_0},
\end{equation}

The general solution of equations (\ref{trapBVPa})--(\ref{trapBVPc}) is
\numparts
\label{sQ1D}
\begin{eqnarray}
 \wrho_1(x,s|x_0)&=A(s)\cosh\alpha(s) (L+x) ,\ x\in [-L,0]\\
 \wrho_2(x,s|x_0)&=B(s)\e^{-\beta(s)x}+ \overline{G}(x, s|x_0) ,\ x \geq 0,
\end{eqnarray}
\endnumparts
where 
\begin{equation}
\alpha(s)=\sqrt{\frac{s}{D}},\quad \beta(s)=\sqrt{\frac{s+\kappa(s)}{D}},
\end{equation}
 and $\overline{G}$ is a Green's function satisfying
 \numparts
\begin{eqnarray}
\label{nGa1D}
&D\frac{\partial^2\overline{G}(x,s|x_0)}{\partial x^2}  -(s+\kappa(s))\overline{G}(x,s|x_0)  = -\delta(x - x_0), \ x<0,\\ 
&  \overline{G}(0,s|x_0)=0.
 \label{nGb1D}
\end{eqnarray}
\endnumparts
Hence,
\begin{eqnarray}
  \label{Ga}
\fl   \overline{G}(x, s| x_0)
 = \frac{\Theta(x - x_0)g(x, s)\widehat{g}(x_0, s) +\Theta(x_0 - x)g(x_0, s)\widehat{g}(x, s)}{\sqrt{[s+\kappa(s)]D} },
\end{eqnarray}
where $\theta(x)$ is the Heaviside function and
\begin{eqnarray}
   g(x, s) = \e^{-\beta(s)x},\quad \widehat{g}(x, s) =\sinh(\beta(s) x). 
\end{eqnarray}
The unknown coefficients $A,B$ are determined from the matching conditions at $x= 0$, which reduce to
\numparts
\begin{eqnarray}
\label{smatcha}
&A(s)\cosh[\alpha(s) L]=B(s),\\
&-\alpha(s)A(s)\sinh[\alpha(s) L]=\beta(s)B(s) +\partial_x\overline{G}(0,s|x_0).
\label{smatchb}
\end{eqnarray}
\endnumparts
Substituting (\ref{smatcha}) into (\ref{smatchb}) gives
\begin{eqnarray}
\label{DN2}
 \left \{\alpha(s)\tanh[ \alpha(s)L] +\beta(s )\right \} B(s)=\frac{\e^{-\beta(s)x_0}}{D} 
\end{eqnarray}
Hence,
\numparts
\begin{eqnarray}
\label{squida}
  & \wrho_1(x,s|x-0)=\frac{\e^{-\beta(s)x_0}}{\Phi(s)D}\frac{\cosh\alpha(s) (L+x)}{ \cosh\alpha(s) L} ,\quad x\in [-L,0],\\
 & \wrho_2(x,s|x_0)=\frac{\e^{-\beta(s)(x+x_0)}}{\Phi(s)D} +\overline{G}(x,s|x_0),\quad x_0,x \geq 0,
\label{squidb}
\end{eqnarray}
\endnumparts
where
\begin{equation}
\label{Psi}
\Phi(s)\equiv  \alpha(s) \tanh [\alpha(s)L] +\beta(s)  .
\end{equation}

The FPT density for absorption is
\begin{equation}
\calF(x_0,t)=\overline{\gamma}_0 \int_0^{\infty} q(x,t|x_0)dx.
\end{equation}
In Laplace space we have
\begin{eqnarray}
\widetilde{\calF}(x_0,s)&=\frac{\kappa_0\overline{\gamma}_0}{s+\gamma_0+\overline{\gamma}_0} \int_0^{\infty} \wrho_2(x,s|x_0)dx\nonumber \\
&=\frac{\kappa_0\overline{\gamma}_0}{s+\gamma_0+\overline{\gamma}_0}\int_0^{\infty}\bigg [\frac{\e^{-\beta(s)(x+x_0)}}{\Phi(s)D} +\overline{G}(x,s|x_0)\bigg ]dx\nonumber \\
&= \frac{\kappa_0\overline{\gamma}_0}{s+\gamma_0+\overline{\gamma}_0}\bigg [\frac{\e^{-\beta(s) x_0}}{\beta(s)\Phi(s)D}  +\frac{1-\e^{-\beta(s)x_0}}{s+\kappa(s)}\bigg ].
\end{eqnarray}
The term $\int_0^{\infty}\overline{G}(x,s|x_0)dx$ was evaluated by integrating both sides of equation (\ref{nGa1D}) with respect to $x$. Using the identity 
$\calT(x_0)=-\partial_s\widetilde{\calF}(x_0,s)_{s=0}$ we obtain the MFPT for absorption:
\begin{equation}
\label{Ta}
\fl \calT(x_0)=\frac{1}{\pi_b}\left (\frac{1}{\gamma_0+\overline{\gamma}_0}+\frac{1}{\kappa_0}\right )  +\frac{L}{\sqrt{\kappa_0\pi_b D}}\e^{-\sqrt{\kappa_0\pi_b/D}x_0},\quad x_0>0.
\end{equation}
The $L$-independent terms represent the average time accrued due to multiple adsorption/desorption events for the set of sample paths that remain within the trapping domain. The final term on the right-hand side of (\ref{Ta}) is then the contribution from paths that spend time outside the trapping domain, and this depends on $L$ and $x_0$. We now observe a major difference from the trapping boundary problem considered in section 2, namely, there is no longer a simple relationship between $\calT(x_0)$ and the MFPT $T(x_0)$ for adsorption. The latter is obtained by taking $\pi_b=1$ and $\pi_d =0$ in equation (\ref{Ta}):
\begin{equation}
\label{Ta0}
 T(x_0)= \frac{1}{\kappa_0}  +\frac{L}{\sqrt{\kappa_0 D}}\e^{-\sqrt{\kappa_0/D}x_0},\quad x_0>0.
\end{equation}
That is, equation (\ref{MFPT}) does not hold. This result reflects the greater complexity of the renewal equations for a partially reactive trap, as we now demonstrate.

\subsection{Renewal equations}
The renewal equations equivalent to (\ref{1DBVPa})--(\ref{1DBVPd}) are
\numparts
 \begin{eqnarray}
\fl \rho_j(x,t|x_0)&=p_j(x,t|x_0)+\pi_d \int^{\infty}_0 dx' \int_0^td\tau''\int_{\tau'}^td\tau \, \rho_j(x,t-\tau'|x')\phi(\tau-\tau') J(x',\tau'|x_0) , \nonumber \\ \fl  \label{trap1}\\
\fl \calF(x_0,t)&=\pi_b\int_0^td\tau  {\phi}(t-\tau) f(x_0,\tau) \nonumber \\
\fl &\quad +\pi_d \int^{\infty}_0dx'\int_0^td\tau' \int_{\tau'}^t d\tau\, \calF(x',t-\tau) \phi(\tau-\tau') J(x',\tau'|x_0),
\label{trap2}
 \end{eqnarray} 
 \endnumparts
 where $p_j(x,t|x_0)$, $j=1,2$, are the probability densities in $\Omega\backslash \calU$ and $\calU$ in the absence of desorption ($\gamma_0=0$), $J(x,t|x_0)$ is the corresponding adsorption probability flux density at the point $x\in [0,\infty)$, and $f(x_0,t)$ is the FPT density for adsorption:
\begin{eqnarray}
f(x_0,t)=\int_0^{\infty}J(x,t|x_0)dx,\quad J(x,t|x_0)=\kappa_0p_2(x,t|x_0).
\end{eqnarray}
 Comparison of equations (\ref{trap1}) and (\ref{trap2}) with the corresponding renewal equations (\ref{ren1}) and (\ref{ren2}) for a partially reactive point-like boundary shows that the former involve an additional integration with respect to the spatial variable  $x' \in \calU$, which considerably complicates the 1D analysis.

Laplace transforming equations (\ref{trap1}) and (\ref{trap2}) with respect to time and using the convolution theorem yields (for $x_0>0$)
\numparts
  \begin{eqnarray}
\fl \wrho_j(x,s|x_0)&=\widetilde{p}_j(x,s|x_0)+\pi_d \widetilde{\phi}(s)\int^{\infty}_0 dx'  \,\wrho_j(x,s|x')\J(x',s|x_0 ) , 
\label{pLTtrap2}
  \end{eqnarray}
  and
 \begin{equation}
 \widetilde{\calF}(x_0,s)= \pi_b \wphi(s)\f(x_0,s)+\pi_d \wphi(s) \int^{\infty}_0 \widetilde{\calF}(x,s)  \J(x,s|x_0)dx.
 \label{LTtrap2}
 \end{equation}
 \endnumparts
 Equations (\ref{pLTtrap2}) and (\ref{LTtrap2}) are examples of Fredholm integral equations. One method for analysing such equations is to perform a Neumann series expansion. In the case of the renewal equation for the FPT density, we have
 \begin{eqnarray}
\label{Neum1D}
\fl  \widetilde{\calF}(x_0,s)&= \pi_b \wphi(s)\bigg [\f(x_0,s)+\pi_d \wphi(s) \int^{\infty}_0 dx\, \f(x,s)  \J(x,s|x_0)\nonumber \\
\fl &\hspace{1cm} +[\pi_d \wphi(s)]^2\int_{0}^{\infty} dx\int^{ \infty}_0dx'  \f(x,s)  \J(x,s|x') \J(x,s|x') +\ldots \bigg ],
     \end{eqnarray}
 which can be rewritten as
    \begin{eqnarray}
\label{Neum1Da}
\fl  \widetilde{\calF}(x_0,s)&= \pi_b \wphi(s)\bigg [\f(x_0,s)+\pi_d \wphi(s) \int^{\infty}_0 dx\, \f(x,s)  \overline{J}(x,s|x_0)\bigg ],
     \end{eqnarray}  
     with
     \begin{equation}
     \label{intJ}
    \overline{J}(x,s|x_0)=\J(x,s|x_0)+\pi_d \wphi(s)\int_{0}^{\infty} dx  \overline{J}(x,s|x') \J(x',s|x_0) .
     \end{equation}
      We know that the series expansion in equation (\ref{Neum1D}) is convergent, since there exists an exact solution of the corresponding equations (\ref{1DBVPa})-(\ref{1DBVPd}). An interesting issue is determining how many terms in the Neumann series are needed to obtain a solution at a desired level of accuracy. 
      
 \begin{figure}[t!]
\centering
\includegraphics[width=12.5cm]{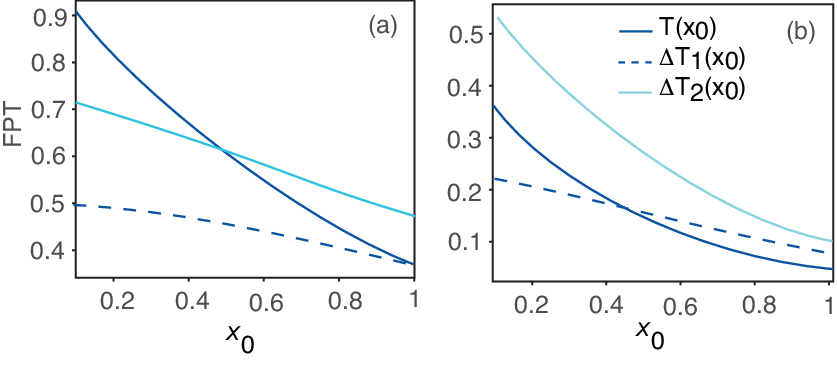} 
\caption{First three terms in the Neumann series expansion of the MFPT ${\calT}(x_0)$ for diffusion in the domain $\Omega=[-L,\infty)$ with a semi-infinite partially reactive trap $\calU=[0,\infty)$. The boundary $x=-L$ is totally reflecting. (a) $\kappa_0=1$, (b) $\kappa_0=5$.}
\label{fig10}
\end{figure}

\begin{figure}[b!]
\centering
\includegraphics[width=11cm]{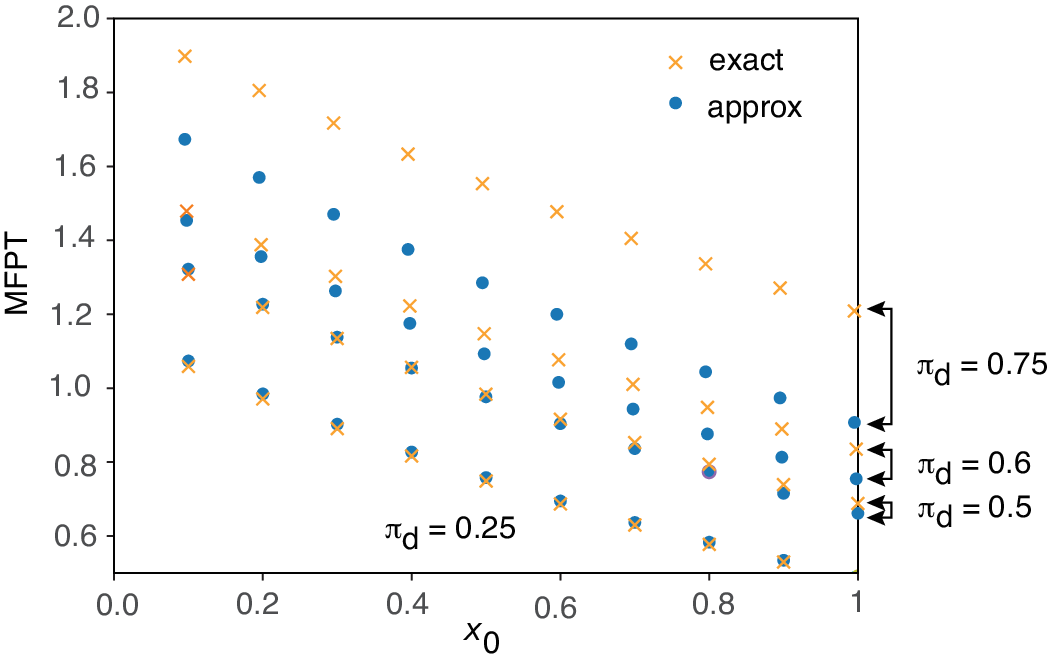} 
\caption{Comparison of $\calT_{\rm app}(x_0)=T(x_0)+\pi_d\Delta T_1(x_0)+\pi_d^2 \Delta T_2(x_0)$ with the exact solution $\calT(x_0)-\langle\tau\rangle/
\pi_b$ for $\kappa_0=1$ and various values of $\pi_d=1-\pi_b$.}
\label{fig11}
\end{figure}

 In order to explore this issue further, consider the MFPT $\calT(x_0)$. Within the renewal approach, $\calT(x_0)$ is related to $T(x_0)$ via an integral equation, which can be obtained directly by performing a small-$s$ expansion of equation (\ref{LTtrap2}). That is, 
substituting the small-$s$ expansions
\begin{equation}
\fl  \widetilde{\calF}(x_0,s)\sim 1-s\calT(x_0),\quad \wphi(s)\sim 1-s\langle \tau\rangle ,\quad \J(x,s|x_0) \sim j_0(x,x_0) -sj_1(x,x_0)
 \end{equation}
into (\ref{LTtrap2}) and collecting all $O(s)$ terms yields the integral equation
 \begin{equation}
  {\calT}(x_0)=  \langle \tau\rangle +T(x_0)+\pi_d\int_0^{\infty} \calT(x)j_0(x,x_0)dx .
  \label{bow}
   \end{equation}
  Setting $\pi_b=1$ in equation (\ref{squidb}) and taking the limit $s\rightarrow 0$ shows that
\begin{eqnarray}
 & j_0(x,x_0)=\kappa_0\bigg [\frac{\e^{-\sqrt{\kappa_0/D})(x+x_0)}}{\sqrt{\kappa_0 D}} +\overline{G}(x,0|x_0)\bigg ],\quad x_0,x \geq 0,
\end{eqnarray}
Iterating equation (\ref{bow}) then gives
\begin{eqnarray}
 \fl &{\calT}(x_0)\nonumber \\
\fl  &=\frac{\langle\tau\rangle}{\pi_b}+\int_0^{\infty} T(x)\bigg [\delta(x-x_0)+\pi_d j_0(x,x_0)+\pi_d^2 \int_0^{\infty}j_0(x,x')j_0(x',x_0)dx' +\ldots \bigg]\nonumber \\
 \fl &=\frac{\langle\tau\rangle}{\pi_b}+T(x_0)+\pi_d \int_0^{\infty} T(x)\overline{j}_0(x,x_0)dx,
 \label{toto}
 \end{eqnarray}
 where $\langle \tau\rangle = 1/\overline{\gamma}_0$ and 
 \begin{equation}
 \overline{j}_0(x,x_0)= {j}_0(x,x_0)+\pi_d\int_0^{\infty}\overline{j}_0(x,x')j_0(x',x_0)dx'.
 \end{equation}
 In Fig. \ref{fig10} we plot $T(x_0)$ together with the higher-order terms
\begin{eqnarray}
\fl  \Delta T_1(x_0)=  \int_0^{\infty} T_0(x)j_0(x,x_0)dx,\quad  
 \Delta T_2(x_0)=  \int_0^{\infty} \Delta T_1(x) {j}_0(x,x_0)dx.
\end{eqnarray}
 It turns out that if $\pi_d< 0.5$ and $\kappa_0=1$, then keeping only the first three terms in the Neumann series expansion yields good agreement with the exact solution (\ref{Ta}). This is illustrated in Fig. \ref{fig11} where we compare $\calT_{\rm app}(x_0)=T_0(x_0)+\pi_d \Delta T_1(x_0)+\pi_d^2\Delta T_2(x_0)$ with the exact expression $\calT(x_0)$ (after dropping the term $\langle\tau\rangle/\pi_b=1/\overline{\gamma}_0$). More generally, the number of terms needed for an accurate truncation of the Neumann series will depend on $\pi_d$ and $\kappa_0$.

 In anticipation of the spectral theory developed in section 5, we end by briefly discussing an alternative version of the integral equation (\ref{intJ}), which can be obtained using sine Fourier transforms. For any $L^2[0,\infty)$ function $F(x)$ set
\begin{equation}
F(x)=\int_0^{\infty}\sin( kx) \widehat{F}(k)dk,\quad \widehat{F}(k) =\frac{2}{\pi}\int_0^{\infty} \sin( kx) F(x)dx.
\end{equation}
We also have
\begin{equation}
\fl \frac{2}{\pi} \int_0^{\infty} \sin(kx)\sin(k'x)dx=\delta(k-k'),\quad \frac{2}{\pi} \int_0^{\infty} \sin(kx)\sin(kx')dk=\delta(x-x').
\end{equation}
Fourier transforming equation (\ref{nGa1D}) shows that
\begin{equation}
\overline{G}(x,s|x_0)=\frac{2}{\pi}\int_0^{\infty}\frac{\sin(kx)\sin(kx_0)}{Dk^2+s+\kappa_0}dk.
\end{equation}
Moreover,
\begin{eqnarray}
\label{id2}
\int_0^{\infty}\sin(kx) \e^{\alpha x}dx=\mbox{Im}\bigg  [\int_0^{\infty}\e^{ikx}\e^{\alpha x}dx\bigg ]=\frac{k}{\alpha^2+k^2}.
\end{eqnarray}
Hence, we can set
\begin{eqnarray}
  \J(x,s|x')=\frac{2\kappa_0}{\pi}\int_0^{\infty}dk \int_0^{\infty} dk' \, \sin(kx)\chi(k,k';s)\sin (k'x'),
\end{eqnarray}
where
\begin{eqnarray}
\fl \chi(k,k';s)= \bigg [  \frac{\delta(k-k')}{Dk^2+s+\kappa_0} +\frac{k}{Dk^2+s+\kappa_0}\frac{k'}{D{k'}^2+s+\kappa_0}\frac{2 D}{\pi\Phi(s)} \bigg ].
\end{eqnarray}
Similarly defining
\begin{eqnarray}
  \J(x,s|x')=\frac{2\kappa_0}{\pi}\int_0^{\infty}dk \int_0^{\infty} dk' \, \sin(kx)\overline{\chi}(k,k';s)\sin (k'x'),
\end{eqnarray}
and substituting into equation (\ref{intJ}) we obtain the alternative integral equation
 \begin{equation}
 \label{chicon}
\overline{\chi}(k,k';s)=\chi(k,k';s)  +  \pi_d \wphi(s) \int_0^{\infty} dk_1 \overline{\chi}(k,k_1;s) \chi(k_1,k';s) .
     \end{equation}
 
 \subsection{Non-Markovian adsorption}
 
Given the relative complexity of the renewal approach for partially absorbing traps, one could argue that it is preferable to work with the original BVP. However, this option is not available in the case of an encounter-based formulation of non-Markovian adsorption. In the absence of desorption, this proceeds along analogous lines to a partially reactive surface \cite{Bressloff22,Bressloff22a}. Instead of the boundary local time, the contact time between a particle and the substrate is given by the occupation time 
\begin{equation}
A(t)=\int_0^t{\bf 1}_{\calU}(X(\tau))d\tau,
\end{equation}
where ${\bf 1}_{\calU}(x)=1$ if $x \in \calU$ and is zero otherwise.
The stopping condition for adsorption becomes 
 \begin{equation}
\label{expA}
{\mathcal T}=\inf\{t>0:A(t)>\widehat{a}\},\quad \P[\widehat{a}>a]\equiv \Psi(a) .
\end{equation}
Define the occupation time propagator as
\begin{equation}
P(x,at|x_0)=\bigg \langle \delta(x-X(t))\delta(a-A(t))\bigg \rangle_{X(0)=x_0},
\end{equation}
where $\langle\cdot \rangle$ denotes the expectation with respect to all sample paths $X(t)$ that satisfy standard Brownian motion within $\Omega$.
Introduce the set of densities
\numparts
\begin{eqnarray}
\fl  p^{\Psi}(x,t|x_0)&= \bigg \langle \delta(x-X(t)\Psi(A(t))\bigg \rangle_{X(0)=x_0}
=\int_0^{\infty}\Psi(a) P(x,a,t|x_0)da,\ x\in [-L,\infty) \nonumber \\\fl \\
\fl \nu^{\psi}(x,t|x_0)&=\bigg \langle \delta(x-X(t)\psi(A(t))\bigg \rangle_{X(0)=x_0}  =\int_0^{\infty}\psi(a) P(x,a,t|x_0)da,\ x >0.
\end{eqnarray}
\endnumparts
with $\psi(a) =-\Psi'a)$ the probability density for the occupation time threshold.
it can be proved that $p^{\Psi}$ and $\nu^{\Psi}$ are related according to the equations \cite{Bressloff22,Bressloff22a}
\begin{equation}
\frac{\partial p^{\Psi}(x,t)}{\partial t}=D\frac{\partial^2 p^{\Psi}(x,t)}{\partial x^2} - \nu^{\psi}(x,t|x_0)  {\bf 1}_{\calU}(x),\quad x\in \Omega  .
\label{nob}
\end{equation}
If $\Psi(a)=1$ for all $a$ then $\psi(a)=0$ and we recover the standard 1D diffusion equation without a trapping region. For almost all other choices for $\Psi$, equation (\ref{nob}) does not yield a closed PDE for $p^{\Psi}(x,t)$ due to the dependence on the absorption flux density $\nu^{\psi}(x,t)$ within the trapping region. However, 
following Refs. \cite{Bressloff22,Bressloff22a} we note that for an exponential distribution $\Psi(a)=\e^{-\kappa_0 a}$, we recover the classical inhomogeneous diffusion equation for a trap with a constant rate of absorption $\kappa_0$:
\begin{equation}
\frac{\partial p(x,t|x_0)}{\partial t}=D\frac{\partial^2 p(x,t|x_0)}{\partial x^2} -\kappa_0 p(x,t|x_0) {\bf 1}_{\calU}(x) , \quad x\in [-L,\infty) .
\label{Robin0}
\end{equation}
Hence, the solution of the BVP with a constant rate of adsorption is equivalent to the Laplace transform of the occupation time propagator with respect to $a$:
\begin{equation}
\calP(x,z,t|x_0)\equiv \int_0^{\infty} \e^{-z a} P(x,a,t|x_0)da=p(x,t|x_0)_{\kappa_0=z}.
\end{equation}
Assuming that the Laplace transform $\calP(x,z,t)$ can be inverted with respect to $z$, the solution for a general distribution $\Psi$ is
 \begin{eqnarray}
  \label{paa}
  p^{\Psi}(x,t|x_0) &= \int_0^{\infty} \Psi(a){\mathbb L}_{a}^{-1}\calP(x,z,t|x_0)da .
  \end{eqnarray}
  The corresponding FPT density is
  \begin{eqnarray}
  \label{faa}
 \fl  f^{\Psi}(x_0,t):&=\int_0^{\infty} \nu^{\psi}(x,t|x_0)  dx
 = \int_0^{\infty}  \bigg [\int_0^{\infty} \psi(a){\mathbb L}_{a}^{-1}\calP(a,z,t|x_0)da\bigg ]dx.
  \end{eqnarray}
 It immediately follows from equation (\ref{Ta0}) that the MFPT for adsorption in the case of a general occupation time threshold distribution $\Psi(a)$ is
 \begin{eqnarray}
T^{\psi}(x_0)&=\int_0^{\infty} \Psi(a){\mathbb L}_{a}^{-1} \bigg [\frac{1}{z}+\frac{L}{\sqrt{zD}}\e^{-\sqrt{z/D}x_0}\bigg ]da \nonumber \\
&=\int_0^{\infty} \Psi(a)\bigg [1+\frac{L}{\sqrt{\pi a D}}\exp\left (-x_0^2/4Da\right  )\bigg ]da.
\label{Tapsi}
\end{eqnarray}
The MFPT for absorption then satisfies an encounter-based version of the integral equation (\ref{bow}):
 \begin{equation}
  {\calT}^{\Psi}(x_0)=  \langle \tau\rangle +T^{\Psi}(x_0)+\pi_d\int_0^{\infty} \calT^{\Psi}(x)j^{\Psi}_0(x,x_0)dx ,
  \label{bowa}
   \end{equation}
   where
   \begin{equation}
  j^{\Psi}_0(x,x_0)=\lim_{s\rightarrow 0} \nu^{\psi}(x,s|x_0).
  \end{equation}

 \setcounter{equation}{0}
\section{Higher-dimensional renewal equations and spectral decompositions.}
In higher spatial dimensions, the surface $\partial \calU$ is spatially extended and this considerably complicates the analysis. First, 
the higher dimensional versions of equations (\ref{ren1}) and (\ref{ren2}) for a partially reactive surface $\partial \calU$ involve a spatial integration with respect to points $\x'\in \partial \calU$. These integrals can be handled using the spectral theory of Dirichlet-to-Neumann operators along the lines of Ref. \cite{Grebenkov23}. Such spectral methods can also be used to handle the matching conditions between $p_1$ and $p_2$ across the spatially extended interface $\partial \calU$ of a partially reactive trap \cite{Bressloff22a}. However, one now has to deal with the fact that adsorption occurs over the separate domain $\calU$, which requires an additional spectral decomposition involving the Laplacian operator within $\calU$. In this section we use spectral theory to derive general expressions for the FPT density $\calF(\x_0)$ of a partially reactive surface $\partial \calU$ and a partially reactive trap $\calU$. The practical application and numerical implementation of these general expressions will be explored elsewhere.

\subsection{Partially reactive surface $\partial \calU $} Let us return to the higher-dimensional configuration shown in Fg. \ref{fig1}(a) for a target surface $\partial \calU$ with  $\calU\subset \Omega \subset \R^d$. The higher-dimensional versions of the renewal equations (\ref{ren1}) and (\ref{ren2}) take the form
\numparts
\begin{eqnarray}
\fl  \rho(\x,t|\x_0)&=p(\x,t|\x_0)+\pi_d\int_{\partial \calU}d\y\int_0^td\tau' \int_{\tau'}^t d\tau\,  \rho(\x,t-\tau|\y)\phi(\tau-\tau') J(\y,\tau'|\x_0),\nonumber \\ \fl \\
\label{2Dren1}
 \fl \calF(\x_0,t)&=\pi_b\int_{\partial \calU}d\y \int_0^td\tau {\phi}(t-\tau') J(\y,\tau'|\x_0) \nonumber
  \\
  \fl &\quad + \pi_d \int_{\partial \calU} d\y\int_0^td\tau' \int_{\tau'}^t d\tau\, \calF(\y,t-\tau)\phi(\tau-\tau') J(\y,\tau'|\x_0).
  \label{2Dren2}
 \end{eqnarray}
 \endnumparts
 Here $p(\x,t|\x_0)$ is the probability density in the absence of desorption and $J(\y,t|\x_0)$ is the corresponding adsorption probability flux into the target at $\y\in \partial \calU$. 
 Laplace transforming the renewal equations using the convolution theorem gives
\numparts
 \begin{eqnarray}
\label{2DLT1}
\wrho(\x,s|\x_0)&=\p(\x,s|\x_0)+\pi_d \wphi(s) \int_{\partial \calU}   \wrho(\x,s|\y) \J(\y,s|\x_0)d\y,\\
\label{2DLT2}
\widetilde{\calF}(\x_0,s)&= \pi_b\wphi(s) \f(\x_0,s)+\pi_d \wphi(s)  \int_{\partial \calU}  \widetilde{\calF}(\y,s) \J(\y,s|\x_0)d\y,
 \end{eqnarray}
 \endnumparts
 where $ \f(\x_0,s)$ is the Laplace transform of the FPT for adsorption:
 \begin{equation}
 f(\x_0,t)=\int_{\partial \calU} J(\y,t|\x_0)d\y.
 \end{equation}
  Equations (\ref{2DLT1}) and (\ref{2DLT2}) are Fredholm integral equations of the second kind. One way to formally solve this type of equation is to use spectral theory \cite{Grebenkov20,Grebenkov23,Bressloff23snob}. For simplicity, we assume a constant rate of adsorption $\kappa_0$ so that the BVP for $\p$ is 
 \numparts
\begin{eqnarray}
\label{PlocLTa}
&D\nabla^2 \p(\x,s|\x_0)-s\p(\x,s|\x_0)=-\delta(\x-\x_0),\ \x \in \Omega\backslash \calU,\\
 & \nabla \p(\x,s|\x_0)\cdot \n=0,\ \x \in \partial \Omega,\\
  &\J(\y,s|\x_0)\equiv -D\nabla \p(\x,s|\x_0) \cdot \n_0=\kappa_0\p(\x,s|\x_0) ,\ \x\in \partial \calU .
  \label{PlocLTc}
\end{eqnarray}
\endnumparts
Here $\n$ is the outward unit normal at a point on $\partial \Omega$ and $\n_0$ is the inward unit normal at a point on $\partial \calU$.
 A well known result from classical PDE theory is that the solution of a general Robin BVP can be computed in terms of the spectrum of a D-to-N (Dirichlet-to-Neumann) operator \cite{Grebenkov20,Grebenkov23}. The basic idea is to replace the Robin boundary condition by the inhomogeneous Dirichlet condition $p(\x,s|\x_0)=h(\x,s)$ for all $\x\in \partial \calU$ and to find the function $h$ for which $\p$ is also the solution to the original BVP. The general solution of the modified BVP is 
 \begin{eqnarray}
 \label{sool}
 \p(\x,s|\x_0)&= \calH(\x,s)+G(\x,s|\x_0),\ \x \in \Omega \backslash \calU, \end{eqnarray}
 where 
 \begin{equation}
  \calH(\x,s)=-D\int_{\partial \calU} \partial_{\sigma'} G(\x',s|\x)h(\x',s)d\x' 
  \end{equation}
  for $\partial_{\sigma'}=\n_0\cdot \nabla_{\x'}$, and $G $ is a modified Helmholtz Green's function:
   \numparts
  \begin{eqnarray}
  \label{nabGa}
 \fl &D\nabla^2 G(\x,s|\x')-sG(\x,s|\x')=-\delta(\x-\x'),\ \x,\x' \in \Omega\backslash \calU,\\
 \fl & G(\x,s|\x') =0,\ \x\in \partial \calU, \quad \nabla G(\x,s|\x')\cdot \n=0,\ \x\in \partial \Omega .
  \label{nabGb}
 \end{eqnarray}
 \endnumparts
The unknown function $h$ is determined by substituting the solutions (\ref{sool}) into equation (\ref{PlocLTc}):
\begin{eqnarray}
\label{fL}
\L_s[h](\x,s)+\frac{\kappa_0}{D}h(\x,s)=-\partial_{\sigma}G(\x,s|\x_0) ,\quad \x \in \partial \calU,
\end{eqnarray}
where $\L_s$ is the D-to-N operator
 \begin{eqnarray}
\label{DtoN}
\L_s[h](\x,s) &=-D\partial_{\sigma}\int_{\partial \calU}\partial_{\sigma'}G(\x',s|\x)h(\x',s)d\x'
\end{eqnarray}
acting on the space $L^2(\partial \calU)$. 

When the surface $\partial \calU$ is bounded, the D-to-N operator $\L_s$ has a discrete spectrum. That is, there exist countable sets of eigenvalues $\mu_n(s)$ and eigenfunctions $v_n(\x,s)$ satisfying (for fixed $s$)
\begin{equation}
\label{eig}
\L_s v_n(\x,s)=\mu_n(s)v_n(\x,s).
\end{equation}
It can be shown that the eigenvalues are non-negative and that the eigenfunctions form a complete orthonormal basis in $L^2(\partial \calU)$. We can now solve equation (\ref{fL}) by introducing an eigenfunction expansion of $h$,
\begin{equation}
\label{eig2}
h(\x,s)=\sum_{m=0}^{\infty}h_m(s) v_m(\x,s).
\end{equation}
Substituting equation (\ref{eig2})
into (\ref{fL}) and taking the inner product with the adjoint eigenfunction $v_n^*(\x,s)$ yields 
\begin{equation}
 \label{spec2}
\p(\x,s|\x_0)=G(\x,s|\x_0)+\frac{1}{D}\sum_{n=0}^{\infty} \frac{{\mathcal V}_n(\x,s){\mathcal V}^*_n(\x_0,s)}{\mu_n(s)+\kappa_0/D},
\end{equation}
with
\begin{equation}
\label{VVn}
{\mathcal V}_n(\x,s)=-D\int_{\partial \calU}v_n(\x',s) \partial_{\sigma'}G(\x',s|\x)d\x'.
\end{equation}
By construction we have the identities
\begin{equation}
\label{ids}
{\mathcal V}_n(\y,s)=v_n(\y),\quad \partial_{\sigma} {\mathcal V}_n(\y,s)=\mu_n(s)v_n(\y,s),\quad \y \in \partial \calU.
\end{equation}
Note that the orthogonality condition 
\begin{equation}
\int_{\partial \calU} v_n^*(\x,s)v_m(\x,s)d\x=\delta_{m,n}
\end{equation}
means that $v_n^*$ and $v_m$ can each be taken to have dimensions of [Length]$^{-(d-1)/2}$.

We can now solve the renewal equations (\ref{2DLT1}) and (\ref{2DLT2}) in terms of the D-to-N eigenfunctions. We focus on the renewal equation for the FPT density. First, it is convenient to rewrite the solution (\ref{spec2}) as \cite{Grebenkov23}
\begin{equation}
\p(\x,s|\x_0)=\p_0(\x,s|\x_0)-\frac{1}{D}\sum_{n=0}^{\infty} \frac{{\mathcal V}_n(\x,s){\mathcal V}^*_n(\x_0,s)}{\mu_n(s)[D\mu_n(s)/\kappa_0+1]},
\end{equation}
with $\partial_{\sigma}\p_0(\y,s|\x_0)=0$ for $\y\in \partial \calU$.
Using the identities (\ref{ids}) we have
 \begin{equation}
\fl  \J(\y,s|\x_0)=-D\partial_{\sigma} \p(\y,s|\x_0) =\frac{\kappa_0}{D}\sum_{n=0}^{\infty} \frac{ v_n(\y,s){\mathcal V}^*_n(\x_0,s)}{\mu_n(s)+\kappa_0/D},\quad \y \in \partial \calU,
 \end{equation}
 and
  \begin{equation}
  \J(\y,s|\y')=\frac{\kappa_0}{D}\sum_{n=0}^{\infty} \frac{ v^*_n(\y',s)v_n(\y,s)}{\mu_n(s)+\kappa_0/D},\quad \y,\y' \in \partial \calU.
 \end{equation}
Following along similar lines to Ref. \cite{Grebenkov23}, we perform a Neumann expansion of the integral equation (\ref{2DLT2}) to give
 \begin{eqnarray}
\label{Neum2D}
 \fl \widetilde{\calF}(\x_0,s)&= \pi_b\wphi(s)\bigg [\f(\x_0,s)+\pi_d \wphi(s) \int_{\partial \calU}\f(\y,s)\J(\y,s|\x_0) d\y+\ldots\bigg ].\nonumber 
     \end{eqnarray}
      Substituting the series expansion of $\J(\y,s|\x_0)$ and using the orthonormality of the D-to-N eigenfunctions implies that $\calF$ is given by a geometric series
 \begin{eqnarray}
 \fl \widetilde{\calF}(\x_0,s)&= \pi_b\wphi(s)  \frac{\kappa_0}{D}\sum_{n=0}^{\infty} \frac{\overline{v}_n(s){\mathcal V}^*_n(\x_0,s)}{\mu_n(s)+\kappa_0/D}\bigg [1
 +\frac{\kappa_0}{D}\frac{\pi_d \wphi(s)}{\mu_n(s)+\kappa_0/D}+\ldots\bigg ],
  \end{eqnarray}
  where
  \begin{equation}
  \overline{v}_n(s)=\int_{\partial \calU} v_n(\y,s)d\y.
  \end{equation}
  We conclude that
     \begin{eqnarray}
     \label{FaF}
 \widetilde{\calF}(\x_0,s)=\sum_{n\geq 0}\left [ \frac{\pi_b  \wphi(s)\lambda_n(s)}{1-\pi_d \wphi(s)  \lambda_n(s)} \right ] \overline{v}_n(s){\mathcal V}^*_n(\x_0,s),  \end{eqnarray}
 with
 \begin{equation} \lambda_n(s)= \frac{\kappa_0}{D}\frac{1}{\mu_n(s)+\kappa_0/D}.
\end{equation}
The FPT density can also be written as
\begin{equation}
  \widetilde{\calF}(\x_0,s)=\frac{\kappa_0}{D}\sum_{n\geq 0} \frac{\pi_b  \wphi(s)\overline{v}_n(s){\mathcal V}^*_n(\x_0,s)}{\kappa_0(1-\pi_d \wphi(s) )/D+ \mu_n(s)} .
\end{equation}

So far we have used the spectrum of the D-to-N operator to solve the multi-dimensional renewal equation (\ref{2DLT2}) for a constant rate of adsorption $\kappa_0$. Another major advantage of this particular eigenfunction expansion is that it is straightforward to extend the analysis to an encounter-based model of adsorption \cite{Grebenkov20,Grebenkov23}. Following section 3, we
treat $z=\kappa_0/D$ as the Laplace variable conjugate to the multi-dimensional version of the local time,
\begin{equation}
\label{loc2D}
\ell(t)=\lim_{\epsilon\rightarrow 0} \frac{D}{\epsilon} \int_0^t\Theta(\epsilon-\mbox{dist}(\X(\tau),\partial \calU))d\tau,
\end{equation}
and set
\begin{equation}
\fl  \J^{\Psi}(\y,s|\x_0)=-D\partial_{\sigma} \p^{\Psi}(\y,s|\x_0)=\int_0^{\infty} \Psi(\ell) {\mathbb L}^{-1}_{\ell} \bigg [z\sum_{n=0}^{\infty} \frac{ v_n(\y,s){\mathcal V}^*_n(\x_0,s)}{\mu_n(s)+z}\bigg ]d\ell,
 \end{equation}
 where $\Psi(\ell)$ is the local time threshold distribution. Assuming that the order of summation and integration can be reversed,
 \begin{eqnarray}
\fl  \J^{\Psi}(\y,s|\x_0)&=\sum_{n=0}^{\infty} v_n(\y,s){\mathcal V}^*_n(\x_0,s) \int_0^{\infty} \Psi(\ell) \bigg [\delta(\ell)-\mu_n(s)\e^{-\mu_n(s)\ell}\bigg ]d\ell\nonumber \\
\fl &=\sum_{n=0}^{\infty} v_n(\y,s){\mathcal V}^*_n(\x_0,s) \widetilde{\psi}(\mu_n(s)).
\end{eqnarray}
We have used the identity $\widetilde{\psi}(s)=1-s\widetilde{\Psi}(s)$. Replacing $ \J(\y,s|\x_0)$ by $\J^{\Psi}(\y,s|\x_0)$ on the right-hand side of equation (\ref{Neum2D}) and summing the resulting geometric series then gives
\begin{equation}
\label{specF}
 \widetilde{\calF}^{\Psi}(\x_0,s)= \sum_{n\geq 0}\left [ \frac{\pi_b  \wphi(s)\widetilde{\psi}(\mu_n(s))}{1-\pi_d \wphi(s)  \widetilde{\psi}(\mu_n(s))}\right ] \overline{v}_n(s){\mathcal V}^*_n(\x_0,s).
\end{equation}

\subsection{Spherically symmetric target.} Note that in the case of the finite interval $[0,L]$ with a partially reactive boundary at $x=0$, the series expansion (\ref{specF}) reduces to equation (\ref{EBLTren2}) with $f^{\Psi}$ given by (\ref{fPsi}). This follows from the fact that the D-to-N operator becomes a scalar multiplier
$\mu_1(s) =\alpha(s)\tanh(\alpha(s)L)$ such that $\overline{v}_1(s)=1$ and $V_1^*(x_0)=\cosh(\alpha(s)(L-x_0))/\cosh(\alpha(s)L)$.
Another geometric configuration where the D-to-N operator reduces to a scalar multiplier is a spherical domain $\Omega =\{\x\in \R^d\,|\, 0\leq  |\x| <R_2\}$ with a spherical target $\calU$ of radius $R_1$ at the centre of $\Omega$ with $R_1<R_2$:
$\calU=\{\x\in \R^d \, |\, 0\leq  |\x|<R_1\}$ and $ \partial \calU=\{\x\in \R^d\, |\, |\x|=R_1\}$.
Suppose that the spherical surface $\partial \calU$ is partially reactive with a constant adsorption rate $\kappa_0$ and waiting time density $\phi(\tau)$ for desorption/absorption. Following \cite{Redner01}, the initial position of the particle is randomly chosen from the surface of the sphere $\calU_0$ of radius $R_0$, $R_1<R_0<R_2$. This allows us to exploit spherical symmetry by writing $\p=\p(R,s|R_0)$ etc. and using spherical polar coordinates. For example, equation (\ref{2DLT1}) reduces to the simpler form
 \begin{eqnarray}
  \label{sprenewal}
  \widetilde{\rho}(R,s|R_0) &= \p(R,s|R_0) +\pi_d \wphi(s) \wrho(R,s|R_1)\f(R_0,s),
  \end{eqnarray}
  where  $\f(R_0,s)$ is the total probability flux into the spherical target:
  \begin{equation}
 \fl  \f(R_0,s)=\Omega_d R_1^{d-1} \J(R_1,s|R_0), \quad \J(R_1,s|R_0)=\kappa_0 \p(R_1,s|R_0),
  \end{equation}
 with $\Omega_d$ the solid angle of the $d$-dimensional sphere. We can identity $\f(R_0,s)$ as the Laplace transform of the FPT for adsorption. Setting $R_0=R_1$ and rearranging determines $\widetilde{\rho}(R,s|R_1)$ and thus
 \begin{equation}
 \label{2DLTrho}
 \widetilde{\rho}(R,s|R_0)= \p(R,s|R_0) + \Lambda(R_1,s|R_0)\p(R,s|R_1),
 \end{equation}
 with
  \begin{equation}
\label{2Dlambo}
\Lambda(R_1,s|R_0)=\frac{\pi_d\  \wphi(s )\f(R_0,s)}{1-  \pi_d  \wphi(s)\f(R_1,s)} .
\end{equation}
Similarly, equation (\ref{2DLT2}) becomes
 \begin{eqnarray}
 \widetilde{\calF}(R_0,s)&=\pi_b  \wphi(s)  \f(R_0,s)  
   +\pi_d  \wphi(s)   \widetilde{\calF}(R_1,s) \f(R_0,s).
 \end{eqnarray}
 Again setting $R_0=R_1$ and rearranging determines $ \widetilde{\calF}(R_1,s)$ such that
 \begin{eqnarray}
 \label{2DLTcalF}
  \widetilde{\calF}(R_0,s)=\frac{\pi_b  \wphi(s )\f(R_0,s)}{1-  \pi_d   \wphi(s)\f(R_1,s)} .
 \end{eqnarray}

The
 probability density $\p(R,s|R_0)$ without desorption satisfies the Robin BVP
\numparts
\begin{eqnarray}
\label{spha}
\fl  &D\frac{\partial^2\p(R,s|R_0)}{\partial R^2} + D\frac{d - 1}{\rho}\frac{\partial \p(R,s|R_0)}{\partial R}-s\p(R,s|R_0) 
   =- \Gamma_d \delta(R-R_0) ,\quad R_1<R<R_2,\nonumber \\ \fl \\
 \fl  &D\frac{\partial \p(R,s|R_0)}{\partial R}=\kappa_0 \p(R,s|R_0) ,\quad R=R_1,\\ \fl &D\frac{\partial \p(R,s|R_0)}{\partial R}=0 ,\quad R=R_2.
\label{sphb}
\end{eqnarray}
\endnumparts
We have set $\Gamma_d=1/(\Omega_dR_0^{d-1})$. Equations of the form (\ref{spha}) can be solved in terms of modified Bessel functions \cite{Redner01}. In particular, one can show that
\begin{eqnarray}
\label{fsp}
 \f(R_0,s)= \frac{\kappa_0/D}{\kappa_0/D+\Lambda(R_1,R_2,s)}\left (\frac{R_0}{R_1}\right )^{\nu} \frac{D_{\nu}(R_0,R_2,s)}{D_{\nu}(R_1,R_2,s)},
\end{eqnarray}
where
\numparts
\begin{eqnarray}
\fl \Lambda(R_1,R_2,s)&=\frac{F_K'(R_1,s)F_I'(R_2,s)-F_I'(R_1,s)F_K'(R_2,s)}{F_K(R_1,s)F_I'(R_2,s)-F_I(R_1,s)F_K'(R_2,s)},\\
 \fl  D_{\nu}(a,b;s)&=I_{\nu}(\alpha(s) a)K_{\nu-1}(\alpha(s) b)+I_{\nu-1}(\alpha(s) b)K_{\nu}(\alpha(s) a),
\end{eqnarray}
and
\begin{equation}
F_I(R,s)=R^\nu I_\nu(\alpha(s) R),\quad F_K(R,s)=R^\nu K_\nu(\alpha(s) R).
\end{equation}
\endnumparts
Here $I_{\nu}$ and $K_{\nu}$ denote the modified Bessel functions of the first and second kind, respectively, with $\nu = 1 - d/2$ and $\alpha(s)=\sqrt{s/D}$.
Substituting (\ref{fsp}) into equation (\ref{2DLTcalF}) and rearranging then gives
\begin{eqnarray}
  \fl \widetilde{\calF}(R_0,s)=\frac{\kappa_0}{D} \frac{\pi_b  \wphi(s) }{\kappa_0(1-\pi_d \wphi(s) )/D+ \Lambda(R_1,R_2,s)} \left (\frac{R_0}{R_1}\right )^{\nu} \frac{D_{\nu}(R_0,R_2,s)}{D_{\nu}(R_1,R_2,s)}.
 \end{eqnarray}
We can immediately identify $\Lambda(R_1,R_2,s)$ as the D-to-N multiplier $\mu_1(s)$. Similarly, for the encounter-based model of non-Markovian adsorption, we have
 \begin{eqnarray}
\fl  \f^{\Psi}(R_0,s)&= \widetilde{\psi}(\Lambda(R_1,R_2,s))  \left (\frac{R_0}{R_1}\right )^{\nu} \frac{D_{\nu}(R_0,R_2,s)}{D_{\nu}(R_1,R_2,s)},
\end{eqnarray}
and
\begin{eqnarray}
  \fl \widetilde{\calF}^{\Psi}(R_0,s)=  \frac{\pi_b  \wphi(s)\widetilde{\psi}(\Lambda(R_1,R_2,s)) }{1-\pi_d \wphi(s) \widetilde{\psi}(\Lambda(R_1,R_2,s))} \left (\frac{R_0}{R_1}\right )^{\nu} \frac{D_{\nu}(R_0,R_2,s)}{D_{\nu}(R_1,R_2,s)}.
 \end{eqnarray}

\subsection{Partially reactive interior trap $\calU$}

In the above analysis we have assumed that the partially reactive target is the surface $\partial \calU$ of a bounded domain $\Omega\backslash \calU$, see Fig. \ref{fig1}. An alternative scenario is that the target interior $\calU$ acts as a partially reactive surface, see Fig. \ref{fig2}. That is, a freely diffusing particle can enter and exit $\calU$, and while it is diffusing within $\calU$, it can be attach to a binding substrate within $\calU$ at a constant rate $\kappa_0$ (adsorption). The particle then subsequently unbinds (desorbs) at a rate $\gamma_0$ or is permanently removed from the surface (absorption) at a rate $\overline{\gamma}_0$. The higher-dimensional BVP analogous to equations (\ref{1Da})--(\ref{1Dc}) takes the form
\numparts
\begin{eqnarray}
\label{1Dtrapa}
 \fl &\frac{\partial \rho(\x,t|\x_0)}{\partial t}=D\nabla ^2\rho(\x,t|\x_0)  - {\bf 1}_{\calU}(\x)\bigg [\kappa_0\rho(\x,t|\x_0)-\gamma_0q(\x,t|\x_0)\bigg ] , \quad \x\in \Omega,\\
\fl &D\nabla  \rho(\y,t|\x_0)\cdot \n =0,\quad \y \in \partial \Omega,
\label{1Dtrapb}
\end{eqnarray}
with $ {\bf 1}_{\calU}(\x)=1$ if $\x\in \calU$ and zero otherwise, and
\begin{equation}
\frac{\partial q(\x,t|\x_0)}{\partial t}=\kappa_0\rho(\x,t|\x_0) -( \gamma_0+\overline{\gamma}_0)q(\x,t|\x_0),\quad \x\in \calU.
\label{1Dtrapc}
\end{equation}
\endnumparts
Here $q(\x,t|\x_0)$ is the probability density that the particle is bound at a point $\x\in \calU$. For simplicity, we assume that the adsorption, desorption and absorption rates are spatially homogeneous. The equivalent renewal equations are
\numparts
 \begin{eqnarray}
\fl \rho(\x,t|\x_0)&=p(\x,t|\x_0)+\pi_d \int_{\calU} d\x' \int_0^td\tau''\int_{\tau'}^td\tau \, \rho(\x,t-\tau'|\x')\phi(\tau-\tau') J(\x',\tau'|\x_0) , \nonumber \\ \fl  \label{tren1}\\
\fl \calF(\x_0,t)&=\pi_b\int_{\calU} d\x\int_0^td\tau  {\phi}(t-\tau) J(\x,\tau|\x_0) \nonumber \\
\fl &\quad +\pi_d \int_{\calU} d\x'\int_0^td\tau' \int_{\tau'}^t d\tau\, \calF(\x',t-\tau) \phi(\tau-\tau') J(\x',\tau'|\x_0),
\label{tren2}
 \end{eqnarray} 
 \endnumparts
 where $p(\x,t|\x_0)$ is the probability density in the absence of desorption ($\gamma_0=0$) and $\J(\x',t|\x_0)$ is the corresponding adsorption probability flux at the point $\x'\in \calU$.
Laplace transforming equations (\ref{tren1}) and (\ref{tren2}) with respect to time and using the convolution theorem yields
\numparts
  \begin{eqnarray}
\wrho(\x,s|\x_0)&=\widetilde{p}(\x,s|\x_0)+\pi_d \widetilde{\phi}(s)\int_{\calU} d\x'  \,\wrho(\x,s|\x')\J(\x',s|\x_0 ) , \label{LTp2}
  \end{eqnarray}
  and
 \begin{equation}
 \widetilde{\calF}(\x_0,s)=  \wphi(s)\pi_b\f(\x_0,s)+ \pi_d\wphi(s)\int_{\calU}\widetilde{\calF}(\x,s)   \J(\x,s|\x_0)d\x.
 \label{LTf2}
 \end{equation}
 \endnumparts
 Here $\f(\x_0,s)$ is the Laplace transform of the FPT for adsorption, defined as
 \begin{equation}
 f(\x_0,t)=\kappa_0\int_{\calU} \p(\x,t|\x_0)d\x.
 \end{equation}
 
 The renewal equations can be formally solved along analogous lines to the case of a partially reactive surface $\partial \calU$ by adapting the spectral decomposition introduced in Ref. \cite{Bressloff22a}. However, as we now show, the analysis is considerably more involved. The first step is to consider the BVP for the Laplace transform $\p(\x,s|\x_0)$. For ease of notation, we denote the solution in the domains $\Omega\backslash \calU$ and $\calU$ by $\p_1$ and $\p_2$, respectively. We then have
 \numparts
\begin{eqnarray}
\label{mzLTa}
& D\nabla^2 \p_1(\x,s|\x_0)-s\p_1(\x,s|\x_0)=-\delta(\x-\x_0),\ \x \in \Omega\backslash \calU , \\
&\nabla \p_1(\x,s|\x_0) \cdot \n=0,\  \x\in \partial \Omega,\\
 &D\nabla^2 \p_2(\x,s|\x_0)-(s+\kappa_0)\p_2(\x,s|\x_0) =0,\, \x\in \calU,
 \label{mzLTc}
 \end{eqnarray}
 supplemented by the continuity conditions
 \begin{equation}
\fl \p_1(\x,s|\x_0)=\p_2(\x,s|\x_0),\quad \partial_{\sigma}\p_1(\x,s|\x_0)=\partial_{\sigma} \p_2(\x,s|\x_0), \ \x \in \partial \calU.
\label{match}
\end{equation}
 \endnumparts
 Following Ref. \cite{Bressloff22a}, we treplace the matching conditions (\ref{match}) by the inhomogeneous Dirichlet condition $\p_1(\x,s|\x_0)=\p_2(\x,s|\x_0)=h(\x,s)$ for all $\x\in \partial \calU$ and then find the function $h$ for which $\p_1$ and $\p_2$ are also the solution to the original BVP. 

The general solution of equations (\ref{mzLTa})--(\ref{mzLTc}) for $\x_0\in \calU$ can be written as
 \numparts
 \begin{eqnarray}
 \label{Usoola}
 \p_1(\x,s|\x_0)&= \calH(\x,s),\ \x \in \Omega \backslash \calU,\\  \p_2(\x,s|\x_0)&= \ocalH(\x,s)+\overline{G}(\x,s|\x_0),\ \x \in \calU,
  \label{Usoolb}
 \end{eqnarray}
 \endnumparts
 where 
 \numparts
 \begin{eqnarray}
  \calH(\x,s)&=-D\int_{\partial \calU} \partial_{\sigma'} G(\x',s|\x)h(\x',s)d\x',\\\ocalH(\x,s)&=D\int_{\partial \calU} \partial_{\sigma'} \overline{G}(\x',s|\x)h(\x',s)d\x',
  \end{eqnarray}
   \endnumparts
$G$ satisfies equations (\ref{nabGa}) and (\ref{nabGb}), and
   \numparts
   \begin{eqnarray}
\fl & D\nabla^2 \overline{G}((\x,s|\x')-(s+\kappa_0)\overline{G}((\x,s|\x')=-\delta(\x-\x'),\ \x,\x' \in \calU , \\
\fl &\overline{G}(\x,s|\x')=0,\ \x \in \partial \calU.
\end{eqnarray}
\endnumparts
The unknown function $h$ is determined by substituting the solutions (\ref{Usoola}) and (\ref{Usoolb}) into the second equation in (\ref{match}):
\begin{eqnarray}
\label{fL2}
\L_s[h](\x,s)=-\overline{\L}_{s+\kappa_0}[h](\x,s)+\partial_{\sigma}\overline{G}(\x,s+\kappa_0|\x_0),\quad \x \in \partial \calU,
\end{eqnarray}
where $\L_s$ is the D-to-N operator (\ref{DtoN}) and
\begin{eqnarray}
\overline{\L}_s[h](\x,s) &=-D\partial_{\sigma}\int_{\partial \calU}\partial_{\sigma'}\overline{G}(\x',s|\x)h(\x',s)d\x'
\label{DtoN2}.
\end{eqnarray}
\endnumparts
Assuming that $\partial \calU$ is bounded, we have the eigenvalue equations
\begin{equation}
\label{eiga}
\L_s v_n(\x,s)=\mu_n(s)v_n(\x,s),\quad \overline{\L}_s \overline{v}_n(\x,s)=\overline{\mu}_n(s)\overline{v}_n(\x,s).
\end{equation}
Since we will ultimately be integrating over the domain $\calU$, we also introduce an eigenfunction expansion of the Green's function $\overline{G}$:
\begin{equation}
\label{oG}
\overline{G}(\x,s|\x_0)=\sum_{j\geq 0} \frac{u_j(\x)u_j(\y)}{s+\kappa_0+\lambda_j},
\end{equation}
where
\begin{equation}
D\nabla^2 u_j(\x)=-\lambda_j u_j(\x),\quad \x \in \calU,\quad u_j(\y)=0,\quad \y \in \partial \calU,
\end{equation}
and 
\begin{equation}
\int_{\calU}u_j(\x)u_k(\x)d\x=\delta_{i,j}.
\end{equation}
The eigenvalues of the negative Laplacian are real and positive definite.

We now solve equation (\ref{fL2}) by introducing an eigenfunction expansion of $h$ with respect to one of the D-to-N operators. For concreteness, we set
\begin{equation}
\label{eigh2}
h(\x,s)=\sum_{m=0}^{\infty}h_m(s) v_m(\x,s).
\end{equation}
Substituting equation (\ref{eigh2})
into (\ref{fL2}) and taking the inner product with the adjoint eigenfunction $v_n^*(\x,s)$ yields the following matrix equation for the coefficients $h_m$:
\begin{equation}
\label{sspec0}
h_n(s) =\frac{1}{D}\overline{\mathcal V}^*_n(\x_0,s)+\sum_{m\geq 1} H_{nm}(s)h_m(s),
\end{equation}
where
 \begin{eqnarray}
\overline{\mathcal V}_n(\x_0,s)&=\frac{D}{\mu_n(s)}\int_{\partial \calU} v_n(\x',s)\partial_{\sigma'}\overline{G}(\x',s|\x)d\x' ,
\end{eqnarray}  
and
\begin{eqnarray}
\fl  H_{nm}(s)&=\frac{D}{\mu_n(s)}\int_{\partial \calU} v_n^*(\x,s)\partial_{\sigma}\left \{\int_{\partial \calU}v_m(\x',s)\partial_{\sigma'}\overline{G}(\x',s|\x) d\x'\right \}d\x .
\label{H}
\end{eqnarray}
Substituting the eigenfunction expansion of the Green'a function, equation (\ref{oG}), we find that
\begin{eqnarray}
 \label{oVn}
\fl \overline{\mathcal V}_n(\x_0,s)=\frac{1}{\mu_n(s)} \sum_{j\geq 0} \frac{v_{n,j}(s)u_j(\x)}{s+\kappa_0+\lambda_j},\quad 
H_{nm}(s)=\frac{1}{\mu_n(s)} \sum_{j\geq 0} \frac{v^*_{n,j}(s)v_{m,j}(s)}{s+\kappa_0+\lambda_j},
\end{eqnarray}
with 
\begin{equation}
 v_{n,j}(s)=D\int_{\partial \calU} v_n(\x',s)\partial_{\sigma'} u_j(\x') d\x' .
 \end{equation}
 
Iterating the implicit matrix equation (\ref{sspec0}) as  Neumann series, we have
\begin{eqnarray}
  \label{hNeum}
\fl &h_n(s) \nonumber \\
\fl &=\frac{1}{D}\bigg [\overline{\mathcal V}_n^*(\x_0,s) +\sum_{m} H_{nm}(s) \overline{\mathcal V}^*_m(\x_0,s)+\sum_{m,m'} H_{nm'}(s)H_{m'm}(s) \overline{\mathcal V}^*_m(\x_0,s)+\ldots\bigg ] \nonumber \\
\fl &=\overline{\mathcal V}_n^*(\x_0,s) +\sum_{m} \overline{H}_{nm}(s) \overline{\mathcal V}^*_m(\x_0,s),
\end{eqnarray}  
with
\begin{equation}
\label{NeumH}
\overline{H}_{nm}(s) ={H}_{nm}(s)+\sum_l \overline{H}_{nl}(s) {H}_{lm}(s).
\end{equation}
Finally, substituting equation (\ref{hNeum}) into equations (\ref{Usoola}) and (\ref{Usoolb}) gives
\numparts
 \begin{eqnarray}
\fl & \p_1(\x,s|\x_0)=\frac{1}{D}\sum_{n,m}{\mathcal V}_n(\x,s)\bigg  [\overline{\mathcal V}_n^*(\x_0,s) +\sum_{m} \overline{H}_{nm}(s) \overline{\mathcal V}^*_m(\x_0,s)\bigg ],\quad \x \in \Omega \backslash \calU,\label{spec1a} \\ 
\label{spec1b}
\fl & \p_2(\x,s|\x_0)\\
\fl &=\overline{G}(\x,s|\x_0)+\frac{1}{D} \sum_{n}\mu_n(s)\overline{\mathcal V}_n(\x,s)\bigg  [\overline{\mathcal V}_n^*(\x_0,s) +\sum_{m} \overline{H}_{nm}(s) \overline{\mathcal V}^*_m(\x_0,s)\bigg ],\  \x \in \calU.\nonumber
 \end{eqnarray}
 \endnumparts

We can now solve the renewal equations (\ref{tren1}) and (\ref{tren2}) in terms of the various eigenfunction expansions. Again we focus on the renewal equation for the FPT density. First,
substituting for ${\overline\mathcal V}_n(\x,s)$ using equation (\ref{oVn}) shows that
 \begin{eqnarray}
\fl & \p_2(\x,s|\x_0)=\overline{G}(\x,s|\x_0)+\sum_{j,k}\Theta_{kj}(s)\frac{u_k(\x)u_j(\x_0)}{(s+\kappa_0+\lambda_k)(s+\kappa_0+\lambda_j)},
\end{eqnarray}
with $\overline{G}$ given by equation (\ref{oG}) and
\begin{eqnarray}
\Theta_{kj}(s)=\frac{1}{D} \sum_{n,m}v_{k,n}(s)\bigg [\delta_{n,m}+\overline{H}_{nm}(s)\bigg ]v^*_{j,m}(s).
 \end{eqnarray}
Since $ \J(\x,s|\x_0)=\kappa_0 \p_2(\x,s|\x_0) $, we obtain the compact expression
 \begin{eqnarray}
  \J(\x,s|\x_0)= \sum_{j,k}u_k(\x) {\chi}_{kj}(s) u_j(\x_0),\quad \x,\x_0\in \calU
  \label{expand}
\end{eqnarray}
with
\begin{equation}
{\chi}_{kj}(s)=\frac{\kappa_0}{s+\kappa_0+\lambda_k}
\bigg [\delta_{k,j}+\frac{\Theta_{kj}(s)}{s+\kappa_0+\lambda_j}\bigg ] .
\end{equation}
Finally, we perform a Neumann expansion of the integral equation (\ref{LTf2}) to give
 \begin{eqnarray}
\label{Neum2Da}
 \fl \widetilde{\calF}(\x_0,s)&= \pi_b\wphi(s)\bigg [\f(\x_0,s)+\pi_d \wphi(s) \int_{ \calU}d\x  \f(\x,s)\J(\x,s|\x_0) +\ldots\bigg ]\nonumber \\
 \fl &=\pi_b\wphi(s)\bigg [\f(\x_0,s)+\pi_d \wphi(s) \int_{\calU}d\x  \f(\x,s)\overline{J}(\x,s|\x_0)\bigg ] ,
     \end{eqnarray}
     where
     \begin{equation}
     \overline{J}(\x,s|\x_0) =\J(\x,s|\x_0) +\pi_d\wphi(s) \int_{\calU} \overline{J}(x,s|\x')\J(\x',s|\x_0)d\x'.
     \end{equation}
     Substututing for $\J$ using equation (\ref{expand}) we find that
     \begin{eqnarray}
  \overline{J}(\x,s|\x_0)= \sum_{j,k}u_k(\x) \overline{\chi}_{kj}(s) u_j(\x_0),\quad \x,\x_0\in \calU
  \label{oexpand}
\end{eqnarray}
where
\begin{equation}
\label{Neumchi}
\overline{\chi}_{kj}(s)=\chi_{kj}(s)+\pi_d\wphi(s) \sum_{l} \overline{\chi}_{kl}(s)\chi_{lj}(s).
\end{equation}
This infinite-dimensional matrix equation is the discrete analog of equation (\ref{chicon}) for a semi-infinite 1D target.

\section{Discussion}  

In this paper, we developed a general probabilistic theory of diffusion-mediated adsorption and absorption at a partially reactive target.
In the case of a target boundary $\partial \calU$, surface reactions were specified by the random local time threshold density $\psi(\ell)$, the waiting time density $\phi(\tau)$ and the splitting probability of desorption $\pi_d$. The density $\psi(\ell)$ determines the typical amount of particle-surface contact time required for an adsorption event to occur, whereas $\phi(\tau)$ determines the typical waiting time before an desorption or absorption (killing) event occurs. If $\pi_d=1$ (no absorption), then the system reduces to the reversible adsorption/desorption framework analysed in Ref. \cite{Grebenkov23}, whereas $\pi_d=0$ represents irreversible adsorption. We also considered the analogous theory for a partially reactive trap $\calU$ in which particle-target contact time corresponds to the amount of time the particle spends within $\calU$ (the so-called occupation time). In both cases, we analysed the stochastic search process in terms of a pair of renewal equations that relate the probability density and FPT density for absorption to the corresponding quantities in the case of irreversible adsorption. The renewal equations effectively sew together successive rounds of adsorption and desorption prior to final absorption of the particle. The advantage of the renewal formulation is that it can be generalised to the case of non-Markovian adsorption using an encounter-based method. We illustrated the theory for a partially reactive surface by considering a Brownian particle in the finite interval $[0,L]$ with a partially reactive boundary at $x=0$ and a totally reflecting boundary at $x=L$. In particular, we investigated the effects of non-Markovian adsorption and desorption on fluctuations of the FPT density for absorption and determined the long-time asymptotics of the FPT density when $\psi(\ell)$ and $\phi(\tau)$ are heavy-tailed. We then considered the example of a semi-infinite partially reactive trap and showed that the solution of the FPT density for absorption takes the form of an infinite Neumann series expansion of a Fredholm integral equation. Finally, we used spectral theory to derive general expressions for the FPT density $\calF(\x_0)$ for partially reactive targets in higher spatial dimensions.

\begin{figure}[t!]
\centering
\includegraphics[height=8cm]{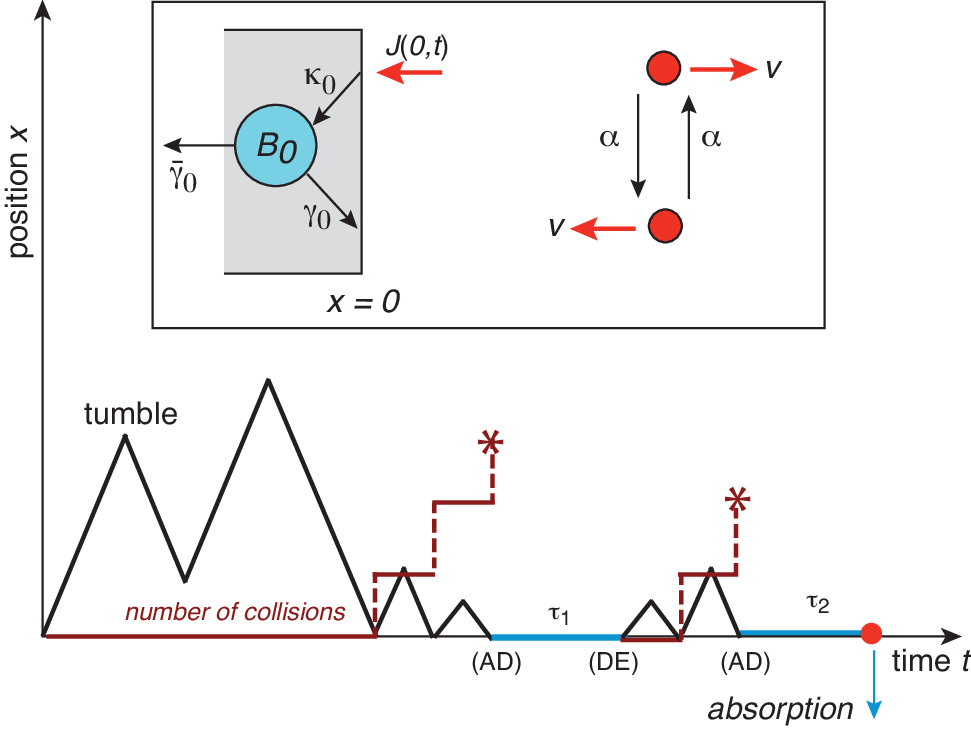}
\caption{Example trajectory of an RTP with a partially reactive wall at $x=0$. When the the number of collisions (shown in red) crosses a randomly generated threshold the RTP is adsorbed (AD) and the collision count is reset to zero. After a waiting time $\tau_1$, the particle desorbs (DE) and the collision count starts increasing again until crossing a new random threshold. Following a second waiting time $\tau_2$, the particle is permanently absorbed. {\bf Inset}: Basic model of an RTP in the case of Markovian adsorption, desorption and absorption. The bound state is denoted by $B_0$.}
\label{fig12}
\end{figure}

There are many possible extensions of the work. First, a more systematic exploration of higher-dimensional geometries beyond target and search domains with spherical symmetry. In terms of the spectral decompositions derived in section 5, this will require truncating the resulting infinite Neumann series. Second, analysing the FPT problem for multiple partially adsorbing/desorbing targets by adapting matched asymptotic and Green's function methods \cite{Bressloff22e,Grebenkov22a}. A third direction is to explore other types of stochastic dynamics in the bulk domain. For example, one could consider a subdiffusive search process and investigate the non-trivial interplay between heavy-tails arising from subdiffusion and those arising from non-Markovian adsorption and desorption \cite{Bressloff23b}. Another example would be to consider the confinement of an active run-and-tumble particles (RTP) by a partially reactive wall.
The analogs of adsorption, desorption and absorption for an RTP are illustrated in the inset of Fig. \ref{fig12}. At the simplest level, run-and-tumble motion in 1D  consists of a particle randomly switching at a constant rate $\alpha$ between two constant velocity states $\pm v$ with $v>0$. Let $p_{k}(x,t)$ be the probability density that at time $t$ particle is at $X(t)=x>0$ and in velocity state $\sigma(t)=k$. The associated evolution equation is given by
\numparts
\begin{eqnarray}
\label{RTPa}
\frac{\partial p_{1}}{\partial t}&=-v \frac{\partial p_{1}}{\partial x}-\alpha p_{1}+\alpha p_{-1},\\
\frac{\partial p_{-1}}{\partial t}&=v \frac{\partial p_{-1}}{\partial x}+\alpha p_{1}-\alpha p_{-1}.
\label{RTPb}
\end{eqnarray}
Suppose that whenever the RTP hits the boundary at $x=0$ it either reflects or binds to the wall at a rate $\kappa_0$ (adsorption). The particle remains in the bound state $B_{0}$ until either desorbing at a rate $\gamma_0$ or being permanently absorbed at a rate $\overline{\gamma}_0$
Let $q(t)$ denote the probability that at time $t$ the particle is in the bound state $B_0$. The boundary condition at $x=0$ takes the form
\numparts
\begin{equation}
\label{RTPc}
J(0,t)\equiv vp_-(0,t)-vp_+(0,t)=\kappa_0 p_-(0,t) -\gamma_0q(t),
\end{equation}
with $q(t)$ evolving according to the equation
\begin{equation}
\label{RTPd}
\frac{dq}{dt}=\kappa_0 p_-(0,t)- (\gamma_0+\overline{\gamma}_0) q(t)
\end{equation}
\endnumparts
Equations (\ref{RTPa})--(\ref{RTPd}) are the analog of the diffusion BVP given by equations
(\ref{1Da})-(\ref{1Dc}). In Ref. \cite{Bressloff22d} an encounter-based model of irreversible adsorption was developed, in which the probability of adsorption depended on the number of particle collisions with the boundary. The renewal framework could be used to incorporate the effects of desorption and absorption by sewing together multiple rounds of bulk RTP motion along analogous lines to Brownian motion. This is illustrated schematically in Fig. \ref{fig12}. We also note that the special case of perfect adsorption ($\kappa_0=v$) combined with Markovian desorption/absorption was previously analysed in Ref. \cite{Angelani17} and subsequently extended to an encounter-based model of non-Markovian absorption in Ref.  \cite{Bressloff23}. In the latter case, the probability of absorption (as distinct from adsorption) was assumed to depend on the amount of time spent in the bound state. 

For both Brownian particles and active RTPs one could also supplement the bulk dynamics with some form of stochastic resetting. At the simplest level, the freely moving particle would instantaneously reset to its initial position $\x_0$ at a sequence of times generated by a Poisson process of constant rate $r$ \cite{Evans11a,Evans11b,Evans13,Evans18}. Moreover, in the case of irreversible adsorption, renewal theory can be used to relate the probability density and FPT density with resetting to the corresponding quantities without resetting. However, care must be taken when extending the renewal theory to allow for the effects of desorption/absorption. Assuming that the particle does not reset when in the bound state, it is necessary to sew together multiple rounds of bulk motion with stochastic resetting, in which the first round starts at the reset point $\x_0$ but subsequent rounds start at a point on the reactive surface $\calU$. That  is, the reset point and initial point are now distinct. Additional complications arise if one also includes an encounter-based model of adsorption \cite{Bressloff22b,Benk22}

Finally, it would be interesting to analyse the effects of permanent absorption on the problem of {\em impatient particles}. This concerns a reaction that is triggered by a threshold
crossing event involving multiple Brownian particles with reversible-binding \cite{Grebenkov17a,Lawley19,Kumar21,Kumar22}. That is, given $N$ independent particles
diffusing in the presence of a reversible  adsorbing surface, when will $K\subset N$ particles be bound to
the target for the first time? The inclusion of absorption would lead to the possibility that the triggering event fails to occur, at least for finite $N$.


\end{document}